\documentclass[twocolumn]{aastex63}

\usepackage{subfigure}
\usepackage{url}
\usepackage{hyperref}
\usepackage{float}
\usepackage{longtable}
\usepackage{natbib}
\usepackage{amsmath}
\usepackage{listings}
\usepackage[normalem]{ulem}
\usepackage{bm}
\usepackage{comment}
\usepackage{array}
\newcolumntype{P}[1]{>{\centering\arraybackslash}p{#1}}
\newcolumntype{M}[1]{>{\centering\arraybackslash}m{#1}}
\usepackage{xcolor, fontawesome}
\definecolor{twitterblue}{RGB}{64,153,255}

\definecolor{Code}{rgb}{0,0,0}
\definecolor{Decorators}{rgb}{0.5,0.5,0.5}
\definecolor{Numbers}{rgb}{0.5,0,0}
\definecolor{MatchingBrackets}{rgb}{0.25,0.5,0.5}
\definecolor{Keywords}{rgb}{1,0,0}
\definecolor{self}{rgb}{0,0,0}
\definecolor{Strings}{rgb}{0,0.63,0}
\definecolor{Comments}{rgb}{0,0.63,1}
\definecolor{Backquotes}{rgb}{0,0,0}
\definecolor{Classname}{rgb}{0,0,0}
\definecolor{FunctionName}{rgb}{0,0,0}
\definecolor{Operators}{rgb}{0,0,0}
\definecolor{Background}{rgb}{0.98,0.98,0.98}
\definecolor{Booleans}{rgb}{0.572,0,0.572}
\definecolor{BuiltinFunction}{rgb}{0.572,0,0.572}
\definecolor{BuiltinConstant}{rgb}{0.572,0,0.572}
\definecolor{Asterisk}{rgb}{0.670,0,1}

\lstdefinelanguage{Python}{
    	numbers=left,
    	numberstyle=\footnotesize,
    	numbersep=7pt,
    	xleftmargin=1.26em,
    	framextopmargin=2em,
    	framexbottommargin=2em,
    	showspaces=false,
    	showtabs=false,
    	showstringspaces=false,
    	frame=l,
    	tabsize=4,
    	stepnumber=1,
	basicstyle=\small\ttfamily,
    	backgroundcolor=\color{Background},
	stringstyle=\ttfamily\color{Strings},
	morekeywords={import,from,class,def,while,if,in,elif,else,not,or,print,break,continue,return,access,as,except,exec,finally,global,import,lambda,pass,print,raise,try,assert},
    	keywordstyle={\color{Keywords}\bfseries}, 
	otherkeywords={[2]*},
	keywordstyle={[2]\color{Asterisk}},
}

\usepackage{color}

\newcommand{\shrug}{\texttt{\raisebox{0.75em}{\char`\_}\char`\\\char`\_\kern-0.5ex(\kern-0.25ex\raisebox{0.25ex}{\rotatebox{45}{\raisebox{-.75ex}"\kern-1.5ex\rotatebox{-90})}}\kern-0.5ex)\kern-0.5ex\char`\_/\raisebox{0.75em}{\char`\_}}}

\urldef\lcourl\url{http://svo2.cab.inta-csic.es/theory/fps/index.php?id=LasCumbres/LasCumbres.SDSS_gp&&mode=browse&gname=LasCumbres&gname2=LasCumbres#filter}

\newcommand{\eg}{{e.g.}}

\newcommand{\tess}{{\it TESS}}

\newcommand{\vsini}{{$V \sin i$}}
\newcommand{\teff}{T$_{\textrm{eff}}$}

\newcommand{\mearth}{{M$_\oplus$}}
\newcommand{\rearth}{{R$_\oplus$}}

\newcommand{\mjup}{{M$_\textrm{Jup}$}}

\newcommand{\eleanor}{\texttt{eleanor}}

\newcommand{\lightkurve}{\texttt{lightkurve}}
\newcommand{\thisstar}{TIC~234284556}
\newcommand{\sig}{$\sigma$\,Ori E}
\newcommand{\PTFO}{PTFO 8-8695}
\newcommand{\Rik}{RIK-210}
\newcommand{\TESS}{\textit{TESS}}

\usepackage{lineno}
\newcommand{\chicago}{Department of Astronomy and Astrophysics, University of
Chicago, 5640 S. Ellis Ave, Chicago, IL 60637, USA}

\newcommand{\nsf}{NSF Graduate Research Fellow}

\newcommand{\unsw}{School of Physics, University of New South Wales, Sydney, NSW 2052, Australia}
\newcommand{\udash}{UNSW Data Science Hub, University of New South Wales, Sydney, NSW 2052, Australia}

\newcommand{\caltech}{Department of Astronomy, MC 249-17, California Institute of Technology, Pasadena, CA 91125, USA}

\newcommand{\austin}{Department of Astronomy, University of Texas at Austin, 2515 Speedway, Stop C1400, Austin, Texas 78712-1205, USA}

\newcommand{\princeton}{Department of Astrophysical Sciences, Princeton University, 4 Ivy Lane, Princeton University, Princeton, NJ 08544, USA}

\begin{document}
\title{Evidence for Centrifugal Breakout around the Young M Dwarf TIC~234284556}
\correspondingauthor{Elsa K. Palumbo}
\email{epalumbo@caltech.edu}

\shorttitle{Centrifugal Breakout around \thisstar} 
\shortauthors{Palumbo et al.}

\author[0000-0001-7967-1795]{Elsa~K.~Palumbo}
\affiliation{\caltech}

\author[0000-0001-7516-8308]{Benjamin~T.~Montet}
\affiliation{\unsw}
\affiliation{\udash}

\author[0000-0002-9464-8101]{Adina~D.~Feinstein}
\altaffiliation{\nsf}
\affiliation{\chicago}

\author[0000-0002-0514-5538]{Luke~G.~Bouma}
\affiliation{\princeton}

\author[0000-0001-8732-6166]{Joel~D.~Hartman}
\affiliation{\princeton}

\author{Lynne~A.~Hillenbrand}
\affiliation{\caltech}

\author[0000-0002-4020-3457]{Michael~A.~Gully-Santiago}
\affiliation{\austin}

\author[0000-0001-5210-1696]{Kirsten~A.~Banks}
\affiliation{\unsw}

\begin{abstract}

Magnetospheric clouds have been proposed as explanations for depth-varying dips in the phased light curves of young, magnetically active stars such as \sig\ and \Rik. However, the stellar theory that first predicted magnetospheric clouds also anticipated an associated mass-balancing mechanism known as centrifugal breakout for which there has been limited empirical evidence. In this paper, we present data from \TESS, LCO, ASAS-SN, and Veloce on the 45 Myr M3.5 star \object{\thisstar}, and propose that it is a candidate for the direct detection of centrifugal breakout. In assessing this hypothesis, we examine the sudden ($\sim$1-day timescale) disappearance of a previously stable ($\sim$1-month timescale) transit-like event. We also interpret the presence of an anomalous brightening event that precedes the disappearance of the signal, analyze rotational amplitudes and optical flaring as a proxy for magnetic activity, and estimate the mass of gas and dust present immediately prior to the potential breakout event. After demonstrating that our spectral and photometric data support a magnetospheric clouds and centrifugal breakout model and disfavor alternate scenarios, we discuss the possibility of a coronal mass ejection (CME) or stellar wind origin of the corotating material and we introduce a reionization mechanism as a potential explanation for more gradual variations in eclipse parameters. Finally, after comparing \thisstar\ with previously identified ``flux-dip" stars, we argue that \thisstar\ may be an archetypal representative of a whole class of young, magnetically active stars.

\end{abstract}


\section{Introduction} \label{sec:intro}

Young ($\lesssim$100 Myr) stars have important implications for planet formation, evolution, and habitability because they tend to be magnetically active \citep[\eg][]{Feigelson_1991, Vidotto_2014} with strong (kG) magnetic fields, their planets tend to be rapidly evolving \citep{mann2020tess}, and their protoplanetary disks may not yet have dissipated \citep[\eg][]{Williams11}. In particular, they can teach us about how, when, and why atmospheric evolution, planetary migration, and other dynamical interactions take place \cite[\eg][]{Rizzuto_2020}. For example, some pre-main sequence stars have X-ray flares that, at their peak luminosity, release more energy in one second than the total X-ray energy from any known solar flare \citep{getman2021xray}. The ionizing radiation from such flares impacts accretion, disk chemistry, and atmospheric erosion \citep{benz10, Waggoner_2019}. Similarly, coronal mass ejections (CMEs)---mass-loss events potentially connected to such flaring activity---may influence radionuclide production in protoplanetary disks, planetary dynamos, and the clearing of debris disks \citep{Osten_2015}.

Young stars exhibit many different classes of photometric variability. One type is the quasi-periodic dimming events of the ``dipper" stars, likely caused by non-uniformly distributed gas and dust in the protoplanetary disk \citep{Bodman_2017,Cody_2010,Ansdell_2016}. Dippers are very common, accounting for 20–30\% of young stellar objects \citep{McGinnis_2015,Ansdell_2019}. Their deep (up to 50\%), often aperiodic eclipse events \citep{Stauffer_2015, Cody_2018} and strong infrared excesses \citep{Hedges_2018} make them easily detectable in time-series photometry and therefore useful for studying disk evolution and planet formation.

More recently, stars without significant infrared excesses but with rapid, periodic photometric variability have been identified. 
\cite{Stauffer_2017, Stauffer_2018, Stauffer_2021} and \cite{Zhan_2019} describe three families of young M dwarfs that exhibit periodic photometric variability that appears to be synchronous with stellar rotation but that does not fit cleanly into previously established categories:

\begin{enumerate}
    \item Scallop shell stars---rapidly-rotating (P$_{\rm rot} < 0.65$ days) M dwarfs whose light curves exhibit multiple wavelike dips that are typically stable over $\sim$75- to 80-day timescales. Such dips have been seen to undergo sudden morphological changes \citep{Stauffer_2017}.
    \item Persistent flux-dip stars---stars with discrete triangularly shaped dips in their light curves. Similarly to scallop shells, dimming events have been seen to suddenly change in depth \citep{Stauffer_2017}.
    \item Transient flux-dip stars---stars with a single prominent and roughly triangular dip. The depth of the dip may vary significantly from cycle to cycle, with more gradual changes sometimes occurring over longer timescales \citep{Stauffer_2017}. \Rik\ is one particularly well-characterized example of such an object \citep{David_2017}.
\end{enumerate}

Work by \cite{gunther_complex_2020} has since shown that the distinction between scallop shells and flux-dip stars may be due only to an observational bias and their quick rotation periods. These authors suggest that a low-cadence sampling rate leads to smearing which makes flux-dip stars look like scallop shells. 

One well-known example of a flux-dip star is the 7-10 Myr old \PTFO, which exhibits 0.45-day-period transit-like dips with shapes, depths, and durations that have varied over a decade of observation and that are synchronous with the host star's rotation \citep{van_Eyken_2012, Yu_2015}. The proposed explanations for the dips for this system have ranged from a precessing Jovian planet \citep{Barnes_2013} to an accretion hotspot \citep{Yu_2015}, or even a small dusty planet \citep{Tanimoto_2020}. More recently, \citet{Bouma20b} have suggested circumstellar material as a likely explanation based on a synthesis of earlier ground-based observations with data from the Transiting Exoplanet Survey Satellite \citep[\tess;][]{Ricker_2014}.

The flux-dip stars also have a number of massive analogs  \cite[e.g][]{Bohlender_2011, Grunhut_2013, Rivinius_2013,Shultz_2020}. Of these, the archetypal example is \sig, a $\approx 1.1$ Myr, B2 star in the Orion complex \citep{Townsend_2013}. \sig's light curve features distinctive double dips with periodicity matching that of the star's rotational period \citep[e.g.][]{Townsend_2013}. Although these dips are morphologically similar to the eclipses of an eclipsing binary \citep{Townsend_2007}, radial velocity data has  excluded a massive ($> 0.04 M_\odot/\sin\ i$) companion \citep{Groote_1977}. 

Theories of the origin of \sig's dips center around magnetic interactions. In particular, \citet{Landstreet_1978} discovered that magnetically trapped plasma could recreate \sig's photometric and spectroscopic variability, with \citet{Townsend_2005} formalizing the underlying theory via a Rigidly Rotating Magnetosphere (RRM) model.

Under this model, \sig's dips are caused by two co-rotating circumstellar plasma clouds originating from stellar winds \citep{Townsend_2005}. Because of \sig's strong magnetic field (7.3 - 7.8 kG at the poles \citep{oksala2015revisiting}) and rapid rotation, the stellar wind would accumulate into relatively dense, stable regions within the star's magnetic field \citep{Owocki_2018}. While the RRM model successfully provided a theoretical basis for \sig's photometric and spectroscopic signatures, \cite{oksala2015revisiting} have since highlighted the need to incorporate additional physics into the RRM model. Regardless of model, magnetospheric clouds are widely accepted for the origin of \sig's variability \citep{Townsend_2005,Oksala_2011, Townsend_2013} and they have also been used to explain the dips of the flux-dip stars \citep{David_2017,Stauffer_2017}. 

Over time, material continues to accumulate in the magnetospheric clouds, but, since  there is a critical point beyond which the magnetic force can no longer contain the magnetospheric cloud's material by balancing the centrifugal force, a mass-balancing mechanism is required \citep{Owocki_2018, Shultz_2020}. 
One such proposed mass-balancing mechanism is centrifugal breakout, in which the ionized gas and dust that make up the magnetospheric clouds accumulate, dragging the magnetic field lines along with them, until the cloud becomes so massive that the magnetic loops constraining the corotating material are stressed and ultimately broken. This centrifugal breakout event would coincide with the previously trapped material being suddenly expelled, with the magnetic field lines reconnecting immediately afterwards \citep[e.g.][]{Townsend_2005}. 

This mechanism has a number of advantages: (1) it can be derived from first principles  \citep[see][]{Townsend_2005}, (2) it is consistent with magnetohydrodynamic simulations \citep{Ud-Doula_2005, Ud-Doula_2008}, and (3) it is supported by stars' observed H$\alpha$ emission \citep{Owocki_2020, Shultz_2020}. However, \citet{Townsend_2013} did not detect any photometric evidence of a breakout event around \sig\ --- nor did the more recent work of \cite{Shultz_2020} in spectroscopic data spanning 20 years around one of \sig's B-type analogs --- leading to to a consideration of alternate mechanisms, such as the diffusion-plus-drift model of \cite{Owocki_2018}. Debate over centrifugal breakout's importance continues today, even though \cite{Owocki_2020} have addressed some of the original objections to centrifugal breakout brought up by \cite{Townsend_2013}.

Here, we consider the light curve of TIC\footnote{\tess\ Input Catalog \citep[TIC;][]{stassun18}} 234284556\footnote{This star is also known as UCAC4~135-177645, 2MASS~J22223966-6303258, WISE~J222239.75-630326.5, DENIS~J222239.6-630325, UPM~J2222-6303, Gaia~EDR3~6405089921141776128,  and APASS 31766662.}, a $\approx$~45-million-year-old, $0.42 \pm 0.02 M_\odot$ star in the Tucana-Horologium association \citep{Kraus_2014} which has co-rotating dip-like features that resemble those of \PTFO, \sig, and the young low-mass stars observed by \citet{Stauffer_2018} and \citet{Zhan_2019}. Notably, we observe a 1.2\%-deep dip that had been present in data from the previous 24 days disappear within a $\sim$1-day interval. We interpret this as evidence for a potential centrifugal breakout event, with more gradual changes in eclipse parameters hinting at a separate role for an additional mass-balancing mechanism. This signal matches observationally with other eclipses that have been attributed to magnetospheric clouds, as well as to the numerical simulations of magnetospheric clouds produced by \citet{Townsend_2008}.

\thisstar\ stands out because it is relatively bright, with $ I = 11.68$ \citep{Denis_2005}, and nearby, at 44.102 $\pm$ 0.028\ pc \citep{Bailer-Jones_2021}, compared to $\approx$130 pc for the younger transient flux-dip stars in the Taurus, Upper Sco, and UCL/LCC clusters. This star has been observed by the \tess\ mission for $\sim$ three sectors over two years. Besides being a promising target for follow-up observations, \thisstar's low mass (0.422 M$_\odot$) also hints at alternatives to \sig's stellar wind mechanism for mass accumulation, such as CMEs.

This paper is organized as follows: In Section \ref{sec:data}, we present our photometric and spectroscopic observations of \thisstar, which include data from \TESS, the All-Sky Automated Survey for Supernovae (ASAS-SN), the Veloce-Rosso spectrograph, and the Las Cumbres Observatory (LCO). In Section \ref{sec:analysis}, we present our analysis of these data and examine depth variations, flare rates, and other relevant features of the light curves. In Sections \ref{sec:dips} and \ref{sec:breakout}, we discuss potential origins of the observed dips and compare our findings with the theoretical predictions of centrifugal breakout. In Section \ref{sec:discussion}, we introduce a mechanism that could be behind more gradual changes in dip size, discuss stellar wind and CME sources of the co-rotating material, and examine our system in the broader context of young stars with similar variability. Finally, Section \ref{sec:conclusion} summarizes our findings, their implications, and the role of future work.

\section{Observations}\label{sec:data}

\subsection{Stellar Parameters}\label{sec:stellar_params}

\thisstar\ is an M3.5 star and a bona fide member of the Tucana-Horologium association (Tuc-Hor), a young moving group with an age of 35-45 Myr \citep{Bell_2015, Crundall19}. Its membership has been previously confirmed based on its proper motion and spectroscopic signatures of youth \citep{Kraus_2014}. Moreover, its short and high-amplitude rotational signal is qualitatively consistent with what would be expected from a young star \cite[\eg][]{Reinhold_2015}. Young, low-mass stars like \thisstar\ tend to have strong magnetic fields, typically ranging from 0.1 to 10 kG \citep{Gregory2012, Shulyak_2019}, and \thisstar's stellar parameters, regular flaring, and rotational signal suggest that it likely is similar.

\begin{figure}[tb]
  \begin{center}
    \includegraphics[width=0.45 \textwidth, trim={0cm 0.0cm 0cm 0cm}, clip=true]{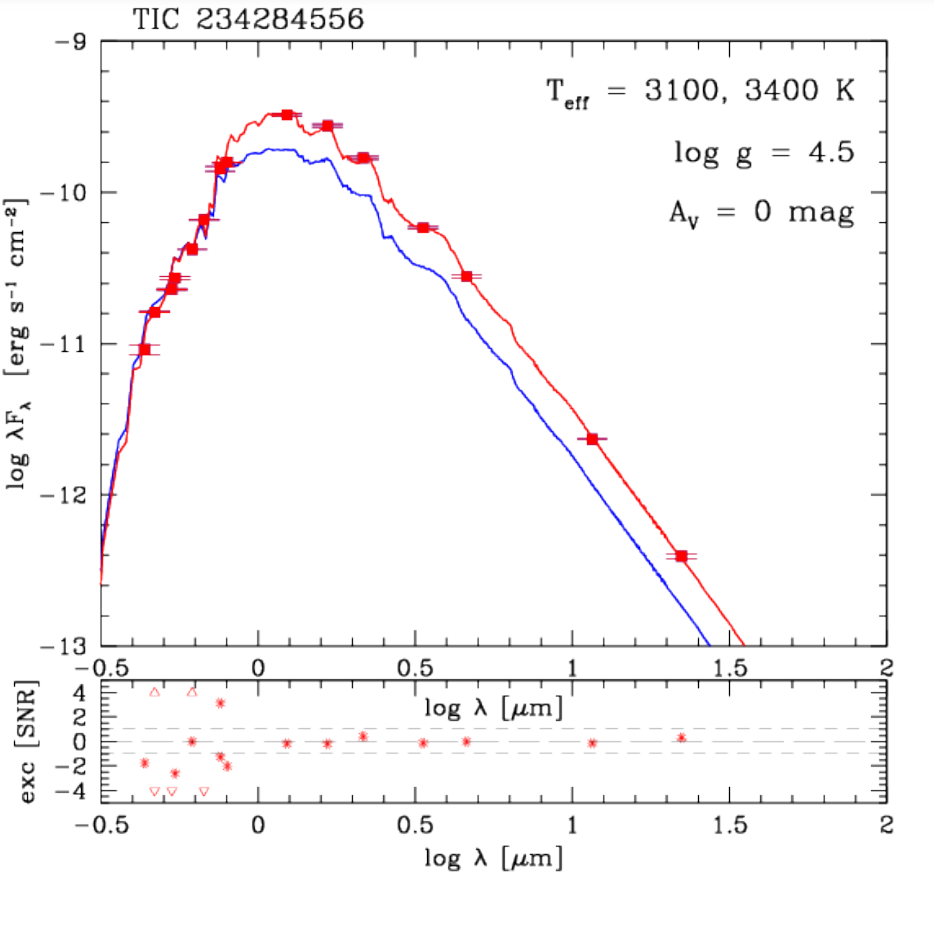}
   \end{center}
  \caption{SED for \thisstar\ along with atmospheric models at the approximate $\pm 1\sigma$ range of the \cite{Stassun_2019} temperature estimate for this star, with the cooler 3100 K model in red with the hotter 3400 K model in blue. For low levels of extinction (A$_{\rm v}$), this plot suggests an effective temperature near the lower limit presented in \citet{Stassun_2019}. Unlike the dipper stars, but like the stars described by \cite{Stauffer_2017, Stauffer_2018, Zhan_2019, Stauffer_2021}, \thisstar\ lacks any detectable infrared excess, indicating that its dimming events are not caused an extended protoplanetary disk.}
    \label{fig:SED}
\end{figure}

Using data from 2MASS, WISE, SDSS, APASS, Gaia EDR3, and Galex (Table \ref{tab:wavelengths}), we constructed the spectral energy distribution (SED) of our target (Figure \ref{fig:SED}), comparing to the NextGen model atmosphere library for stars of similar temperature and metallicity \citep{Hauschildt_1999}. We find the SED is well-described by a $3100 \pm 60$ K model atmosphere with no infrared excess or interstellar extinction; this temperature is consistent with the estimate of $3249 \pm 157$~K provided by \cite{Stassun_2019}. The lack of IR excess indicates that the primordial protoplanetary disk has already dissipated, effectively ruling out a dip-causing mechanism due to a massive, extended disk, as observed in the dipper stars.

The SED also allows us to place a numerical upper limit on the amount of dust present at the Kepler corotation radius $r_K$, the orbital distance where an object's orbital period coincides with the stellar rotational period, which is given by

\begin{equation}\label{eq:Kepler_radius}
r_K = \left(\frac{ G M_*}{4 \pi^2}P_{\rm{rot}}^2\right)^{1/3}
\end{equation}

\noindent where $P_{\rm{rot}}$ is the rotational period of the star, M$_*$ its mass, and $G$ is the gravitational constant.

To estimate the amount of dust that could exist near the corotation radius given the observed SED, we first found the \teff\ of material at the Kepler co-rotation radius to be $T_{\rm K}= 806 \pm 41$ K. This temperature corresponds to emission at wavelengths near the WISE1 bandpass near $3.6\ \mu$m. We therefore used the uncertainty from the WISE1 observation to calculate the increase in flux density at $3.6\ \mu$m that we can attribute to dust. Finally, using stellar parameters from Table \ref{tab:stellarpaarameters} and Equation 1 from \citet{Buemi_2007}, we find our 1$\sigma$ estimate for the dust mass to be

\begin{equation}
    M_{\rm{dust}} = 1.3 \times 10^{-4} M_\odot \bigg(\frac{\chi_\nu}{1\, \rm{cm}^2\, \rm{g}^{-1}}\bigg)^{-1},
\end{equation}

\noindent where $\chi_\nu$ is the opacity of the dust at the observing frequency. For typical opacities ($\chi_\nu$ = 1-10 cm$^2$ g$^{-1}$) this would correspond to a dust mass of $4.2$ to $42$ \mearth.

%$2.5 \times 10^{28}\ g$ to $2.5 \times 10^{29}\ g$. 

\begin{deluxetable*}{r r r}[!ht]
\tabletypesize{\footnotesize}
\tablecaption{Wavelength-Specific Magnitudes for \thisstar\ \label{tab:wavelengths}}
\tablehead{\colhead{Bandpass} & \colhead{Value} & \colhead{Reference}}
\startdata
WISE4 [22.24 $\mu$m] & 8.644 $\pm$ 0.345 & \citet{WISE_2014}\\
WISE3 [11.56 $\mu$m] & 8.867 $\pm$ 0.027 & \citet{WISE_2014}\\
WISE2 [4.600 $\mu$m] & 9.012 $\pm$ 0.020 & \citet{WISE_2014}\\
WISE1 [3.350 $\mu$m] & 9.192 $\pm$ 0.024 & \citet{WISE_2014}\\
H [2.159 $\mu$m] & 9.345 $\pm$ 0.023 & \citet{2MASS_2003}\\
K [1.662 $\mu$m] & 9.588 $\pm$ 0.026 & \citet{2MASS_2003}\\
J [1.235 $\mu$m] & 10.183 $\pm$ 0.024 & \citet{2MASS_2003}\\
RP [0.799 $\mu$m] & 11.9489 $\pm$ 0.0055  & \citet{GaiaDR3}\\
i [0.759 $\mu$m] & 12.489  $\pm$ 0.04 & \citet{Zacharias_2012}\\
G [0.673 $\mu$m] & 13.1820 $\pm$ 0.0030 & \citet{GaiaDR3}\\
r [0.617 $\mu$m] & 14.038 $\pm$ 0.00 & \citet{Zacharias_2012}\\
V [0.544 $\mu$m] & 14.645 $\pm$ 0.02 & \citet{Zacharias_2012}\\
BP [0.532 $\mu$m] & 14.8089 $\pm$ 0.0076 & \citet{GaiaDR3}\\
g [0.469 $\mu$m] &  15.338 $\pm$ 0.01 & \citet{Zacharias_2012}\\
B [0.436 $\mu$m] & 16.245 $\pm$ 0.08 & \citet{Zacharias_2012}\\
\enddata
\end{deluxetable*}

\begin{figure}[tb]
  \begin{center}
  \includegraphics[width=.35\textwidth, trim={0cm 0.0cm 0cm 0cm}, clip=true]{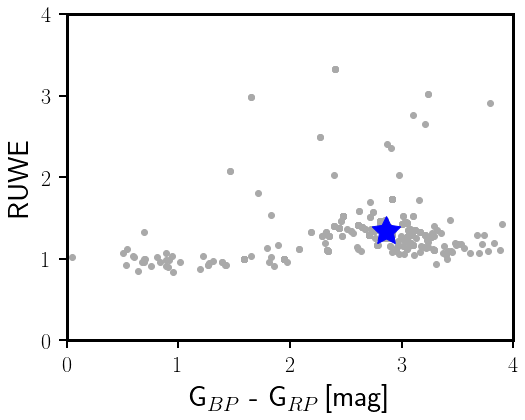}
    \includegraphics[width=.35\textwidth, trim={0cm 0.0cm 0cm 0cm}, clip=true]{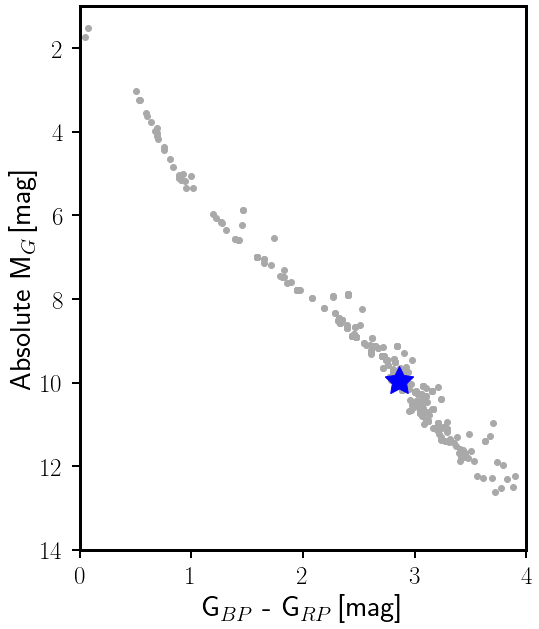}
   \end{center}
  \caption{Evidence for nonbinarity of \thisstar. Point estimates for \thisstar\ (blue star) and all other high confidence Tuc-Hor members (gray circles) are shown, where a difference of Gaia's Rp and Bp bandpasses is used to quantify a star's color, a proxy for spectral type.  Top Panel: Assessing astrometric binarity. Note the color-dependence of typical RUWE values, which casts doubt on the interpretation that \thisstar's RUWE of 1.348 is indicative of it being an astrometric binary. Bottom Panel: Assessing photometric binarity. Note that the point estimate for \thisstar\ is not above the intrinsic scatter, suggesting that our target is not a photometric binary.}\label{fig:binarity}
\end{figure}

\thisstar\ has a Renormalized Unit Weight Error (RUWE) of 1.348 \citep{GaiaDR3}, which might be considered suggestive of astrometric binarity \cite[e.g.][]{Belokurov_2020}. However, Figure \ref{fig:binarity}'s plot of the RUWE of the high-confidence Tuc-Hor members from \citet{gagne18} shows a color dependence, with Tuc-Hor members of a similar color having similar RUWE values, casting doubt on the binary interpretation. For an independent confirmation of this assessment, we plotted the Hertzsprung-Russel diagram for the same list of high-confidence Tuc-Hor members (Figure \ref{fig:binarity}), finding \thisstar\ to be consistent with the expected position for a single star. 

\begin{deluxetable*}{r r r}[!ht]
\tabletypesize{\footnotesize}
\tablecaption{Summary of Stellar Parameters for \thisstar \label{tab:stellarpaarameters}}
\tablehead{\colhead{Parameter} & \colhead{Value} & \colhead{Reference}}
\startdata
TESS Designation & \thisstar & \citet{Stassun_2019}\\
Gaia EDR3 Designation & 6405089921141776128 & \\ 
RA [J2000] & 22h 22m 39.69s & \cite{GaiaDR3}\\
Dec [J2000] & -63$^{\circ}$\ 03'\ 25.83'' & \cite{GaiaDR3}\\
Spectral Type & M3.5 & \citet{Kraus_2014}\\
m$_{TESS}$ & 11.8868 $\pm$ 0.0080 & \citet{Stassun_2019}\\
I$_{\rm mag}$ & 11.68 $\pm$ 0.03 & \citet{Denis_2005}\\
\vsini\ [km $s^{-1}$] & 11.9 $\pm$ 0.4 & This Work\\
P$_{\textrm{rot}}$\ [days] & 1.1066 $\pm$ 0.0003 & This Work\\
Distance [pc] & 44.102 $\pm$ 0.028& \cite{Bailer-Jones_2021}\\
Mass [M$_\odot$] & 0.422 $\pm$ 0.020 & \citet{Stassun_2019} \\
Radius [R$_\odot$] &  0.428 $\pm$ 0.013 & \citet{Stassun_2019} \\ 
\teff\ [K] & 3100 $\pm$ 60 & This work   \\
Age [Myr] & 45 $\pm$ 4 & \citet{Bell_2015}\\
$\rm Log_{10}$(g$_*$) [cm s$^{-2}$] & 4.8014 $\pm$ 0.0051 & \cite{Stassun_2019}\\
r$_K$ [R$_*$] & 7.89 $\pm$ 0.27 & This Work \\ 
i$_{\rm rot}$ [$^\circ$] & 37.5 $\pm$ 2.0 & This Work \\
\enddata
\end{deluxetable*}

% $r_K$ is 7.89+/-0.27 stellar radii; uncertainty propogation in Jupyter notebook 

% using v = (2 pi R_*)/P 
% sin i = v sin i/ v 
% 37.5 $\pm$ 3.6

\begin{figure*}[tb]
  \begin{center}
    \includegraphics[width=\textwidth, trim={0cm 0.0cm 0cm 0cm}, clip=true]
    {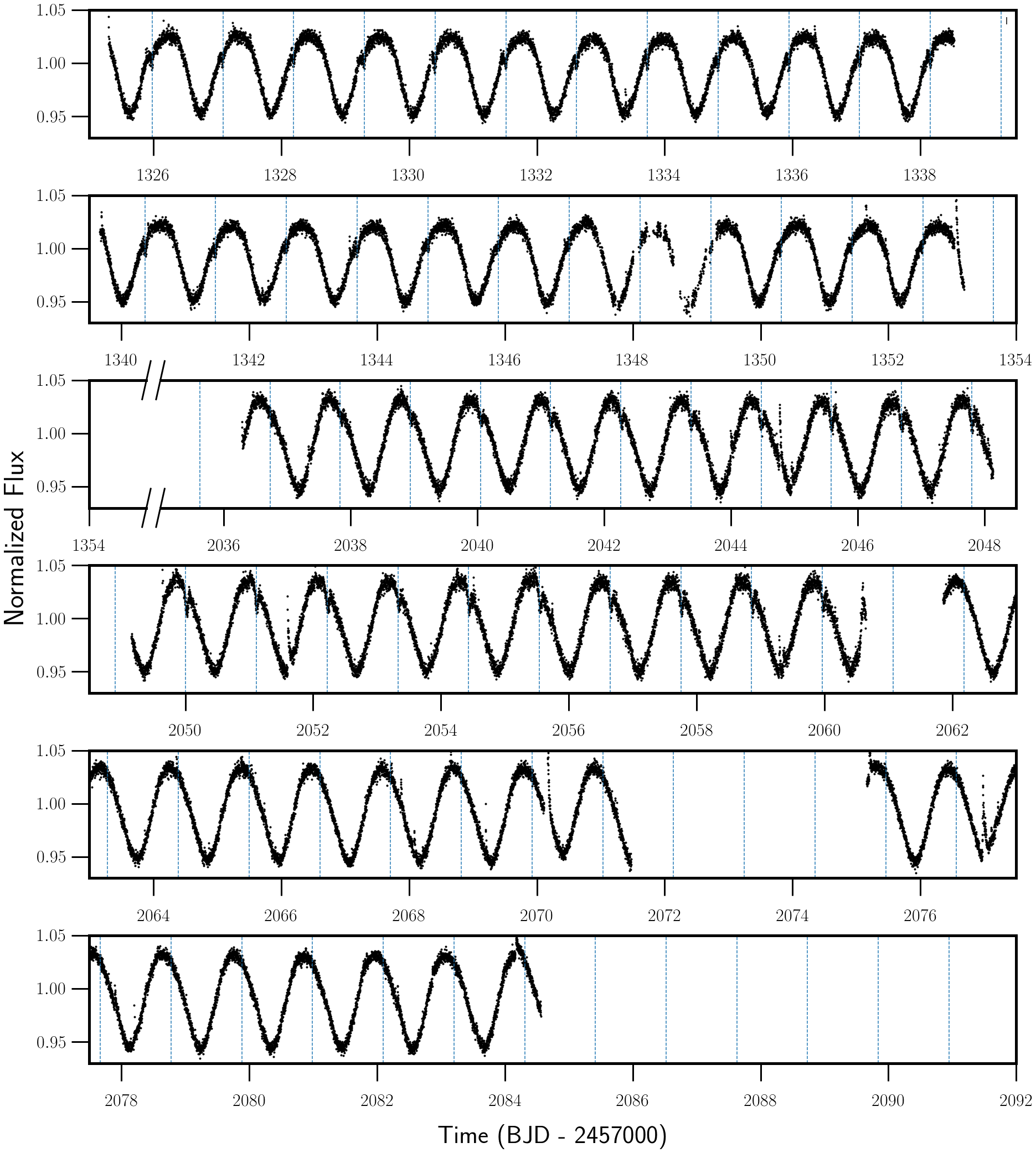}
   \end{center}
  \caption{Two-minute cadence \TESS\ data on \thisstar, with data and errorbars in black. Dashed blue lines indicate the anticipated midpoint of each dimming event, projected forward using the rotational period of the star and the midpoint of the first dip in Sectors 1 and 27. The upper limit of the y axis is held at a normalized flux of 1.05, so not all flares are visible in their entirety. The dip disappears quickly between days 2060 and 2062, consistent with centrifugal breakout.}
    \label{fig:three_sectors}                                                     
\end{figure*}

\subsection{\TESS} \label{sec:TESS}

 \TESS\ \citep{Ricker_2014} observed \thisstar\ in three sectors over two years: Sector 1 (2018 July 25 - 2018 August 22), Sector 27 (2020 July 4 - 2020 July 30), and Sector 28 (2020 July 30 - 2020 August 26). \thisstar\ was pre-selected as a short cadence target through the TESS Guest Investigator program and is also available in the Full-Frame Images (FFIs).\footnote{The target was proposed in program GI G011175, G011266, G011180, G03265 and G03226.}
 
\begin{figure*}[p]
  \begin{center}
    \includegraphics[width=0.9\textwidth, trim={0cm 0.0cm 0cm 0cm}, clip=true]{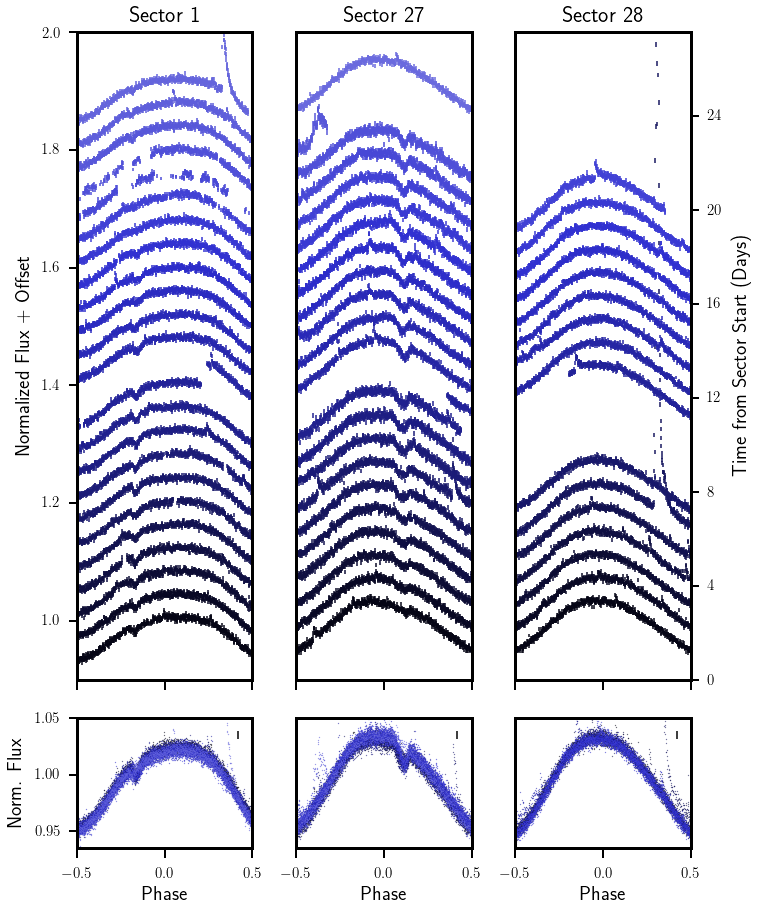}
   \end{center}
  \caption{\tess\ light curve folded on the stellar rotation period. The observed dip evolves throughout Sectors 1 and 27, quickly disappearing between Sectors 27 and 28. An unusually symmetric flare-like brightening occurs at the very end of Sector 27, with a $>$ 120\% flare following some days later in Sector 28.  A +.04 vertical offset is added for each new rotational period, so that time progresses upwards on the plot. To emphasize the short timescales involved, the first rotational period of Sector 28 is plotted in the Sector 27 panel.}\label{fig:stacked_sectors}
\end{figure*}

 As a precaution against data processing effects, we compared results from both short-cadence and FFI light curves throughout this work, where we relied on the open-source Python package \eleanor\ \citep{Feinstein_2019} for extracting light curves from the FFIs. However, for our final analysis, we used all of the two-minute-cadence \tess\ data produced by the NASA Ames' Science Processing and Operations Center (SPOC) pipeline \citep{Jenkins_2016}, only neglecting cadences flagged by that pipeline as having potential quality concerns. The resulting short-cadence \tess\ light curves are shown in Figure \ref{fig:three_sectors}, with phase folded and stacked versions of the same data in Figure \ref{fig:stacked_sectors}. 
 
 In our original data processing, we applied Gaussian Process (GP) regression to the long-cadence data in order to model \thisstar's rotational signal. We used the \texttt{gp.terms.RotationTerm} kernel from the Python package \texttt{exoplanet} \citep{Foreman_Mackey_2017,exoplanet:exoplanet} to model stellar variability as the combined behavior of two underdamped simple harmonic oscillators, one with a period corresponding to the rotational period of the star, and the other with half that period. We then used an iterative approach to define hyperparameters describing the GP, and to identify and mask outliers. 
 
With this GP fit, we removed a model of stellar rotation from the \tess\ short-cadence data. Inspecting the light curve with the rotation removed, we detected a dip with an orbital period of 1.1065 $\pm$ 0.0037 days and a depth of 0.487 $\pm$ 0.043$\%$ over the sector. A closer investigation identified that the dip duration and depth varies gradually over the sector, ranging from 0.12$\%$ to 0.75$\%$ over the course of Sector 1. The overall trend is toward a decreasing dip depth over time, a phenomenon discussed in more detail in Section \ref{sec:mass_loss}. The dip's period is consistent with the 1.1071 $\pm$ 0.0036 day stellar rotation period that we measured from the GP, suggesting a co-rotating object. Moreover, the shape of the event appears to be asymmetric, with a slower egress than ingress.
 
This system was re-observed by \tess\ in Sectors 27 and 28. In Sector 27, a very similar dip is apparent, although with a deeper depth, 0.756 $\pm$ 0.043$\%$ overall, with variation from 0.38$\%$ to 1.2$\%$ over the sector. We also see a change in phase and a reversal of the asymmetry, with the ingress now slower than the egress, a phenomenon discussed in more detail in Section \ref{sec: profiles}. Surprisingly, in Sector 28 there is no detectable periodic dip, even though it is separated from Sector 27 by only 1.2 days.

At a time that coincides with the rapid disappearance of the eclipse and immediately before the end of Sector 27, we see an anomalous brightening event with a morphology unlike the typical steep rise and exponential decay exhibited by a stellar flare (see the last partial orbit of Sector 27 in Figure \ref{fig:stacked_sectors}). Although \tess\ data near the start and end of an orbit can have significant systematics, this signal does not appear to be related to scattered Earthshine. Instead, we see similar behavior across the entire pixel response function, and neighboring stars do not exhibit such a feature, so we infer that the signal appears to be astrophysical and related to \thisstar.
 
\begin{figure*}[ht]
  \begin{center}
    \includegraphics[width=.9\textwidth, trim={0cm 0.0cm 0cm 0cm}, clip=true]{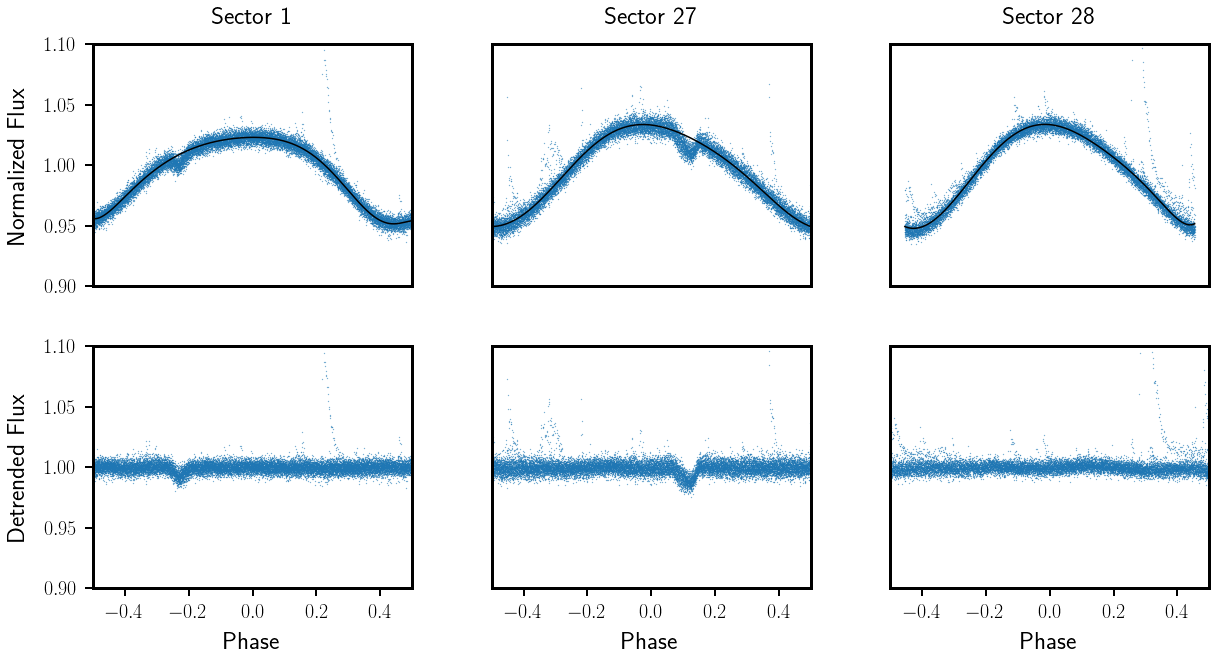}
   \end{center}
  \caption{Detrending \tess\ light curves. Top Panels: The normalized flux folded on the stellar rotation period is shown in blue with a ninth degree polynomial fitted to each sector in black. Bottom Panels: The detrended folded normalized flux, obtained by dividing the light curves from each sector by the corresponding ninth degree polynomial. Note that the amplitude of the variability is larger in Sectors 27 and 28 than in Sector 1, an indicator of heightened magnetic activity around the time of the dip's disappearance.}\label{fig:detrended}
\end{figure*}

\thisstar\ has a strong and roughly sinusoidal rotational signature that is visible in Figures \ref{fig:three_sectors} and \ref{fig:stacked_sectors}. 
In our final data analysis, we took advantage of this stable rotational signal to fit a ninth-degree polynomial to the folded, normalized light curve of the two-minute cadence data with the dimming events masked; the polynomial was created using \texttt{numpy.polyfit}. After repeating this process for each \tess\ sector, we divided our original data by a periodic version of this polynomial fit to obtain our detrended light curves, as shown in Figure \ref{fig:detrended}.
 
 %NASA Ames' Science Processing and Operations Center (SPOC) pipeline \citep{Jenkins_2016} processed the \TESS\ data on \thisstar\ and searched it for periodic transit-like signals. Although the SPOC pipeline identified periodic Sector 1 and 27 dimmings in \thisstar\ as a threshold-crossing event, significant systematics in the detrended data due to the star's strong (7-8\% full amplitude; 1.1 day period) rotational signal precluded an accurate determination of the dimming event parameters and TIC 234284556 was not labeled as a \TESS\ Object of Interest (TOI)\footnote{The SPOC pipeline's data validation reports for TIC 234284556's Sector 1 and Sector 27 data, respectively are available here: \url{https://archive.stsci.edu/missions/tess/tid/s0001/0000/0002/3428/4556/tess2018206190142-s0001-s0001-0000000234284556-01-00106_dvs.pdf},\url{https://archive.stsci.edu/missions/tess/tid/s0027/0000/0002/3428/4556/tess2020187183116-s0027-s0027-0000000234284556-00362_dvm.pdf}}. 
 
 %Our group originally flagged \thisstar\ as having an interesting periodic signal during a systematic search for young planets in the \tess\ FFIs. We included \thisstar\ on a list of potentially young stars based on \banyan, a Bayesian approach to identifying young stars via their kinematics \citep{gagne18}. We originally found the Gaia kinematics to be consistent with Tuc-Hor and thus was included in our original search. 
 
\subsection{ASAS-SN}\label{sec:asas-sn}

\thisstar\ was also observed by ASAS-SN \citep{Shappee_2014, Jayasinghe_2021}, with 208 data points collected over 4.4 years, from 2014 May 12 - 2018 September 24 (HJD 2456789.859 - 2458385.597). There is overlap between the observations of \thisstar\ during \tess\ Sector 1 and the ASAS-SN data. The uncertainties on and times between ASAS-SN's measurements are too large for us to claim a detection of a dip during these additional four years of data. However, these data do confirm the stability of \thisstar's rotational period and phase over timescales of years; in particular, the ASAS-SN database \citep{Jayasinghe_2021} reports a point estimate of 1.1066 days, which is consistent with the value that we measure from \tess. 

\begin{figure}[ht]
  \begin{center}
    \includegraphics[width=0.45\textwidth, trim={0cm 0.0cm 0cm 0cm}, clip=true]{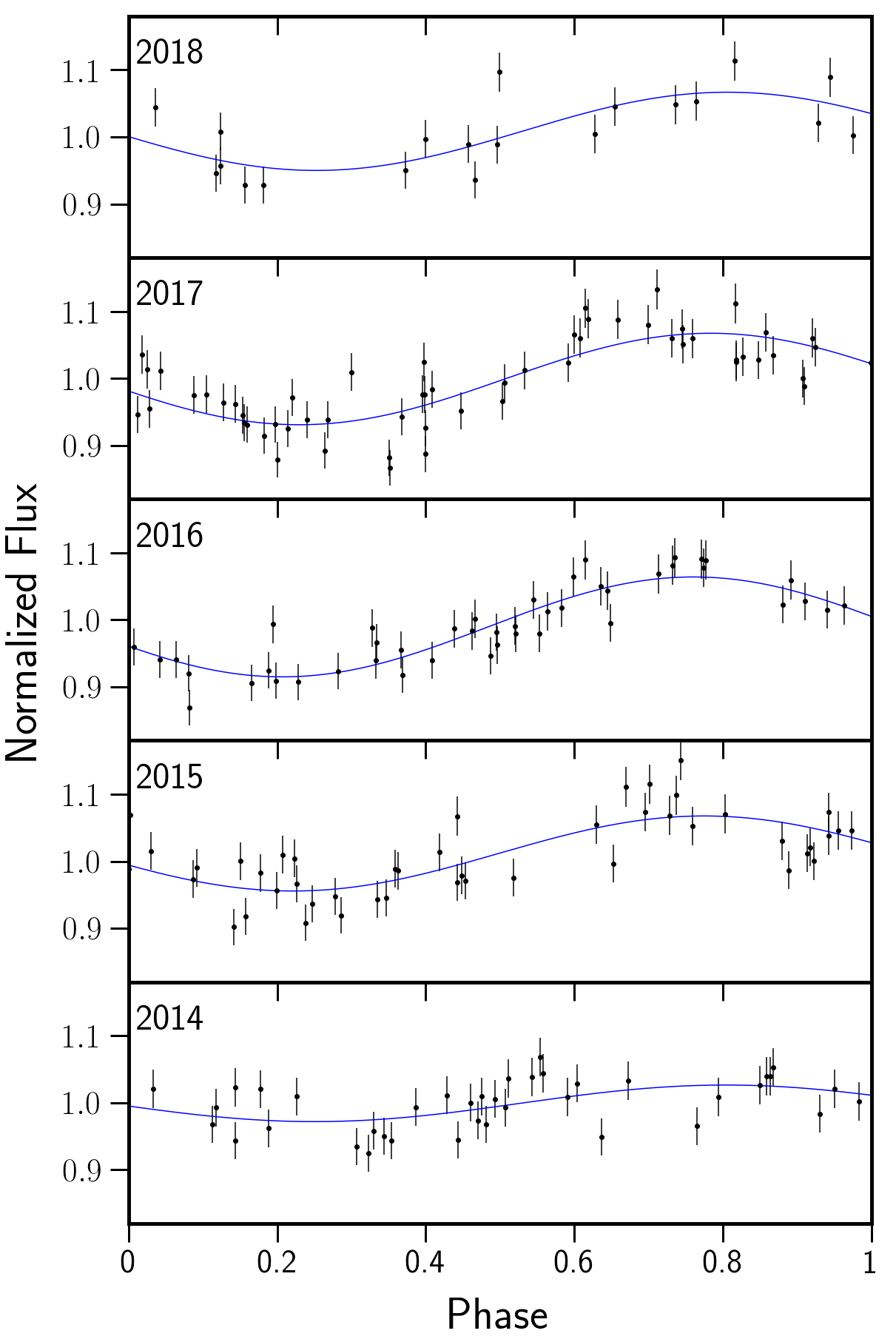}
   \end{center}
  \caption{ASAS-SN data, labeled by year and shown with the best-fit sinusoid in blue. Note the variation in amplitude from year to year, with the semi-amplitude ranging from 2.73 $\pm$ 0.77\% in 2014 to 7.45 $\pm$ 0.60\% in 2016. This variability indicates that the quasi-sinusoidal signal is driven by starspot evolution rather than an alternate mechanism such as Doppler beaming.}
    \label{fig:ASASN}
\end{figure}

While the period and phase are consistent over many years, the amplitude observed by ASAS-SN is not. We fit an individual sine wave with a fixed period but amplitude, phase, and a constant offset as free parameters to the five years of ASAS-SN data and to the three sectors of \tess\ data. We find that the semi-amplitude of the signal in the ASAS-SN data varies from a minimum of 2.73 $\pm$ 0.77\% in 2014 to a maximum of 7.45 $\pm$ 0.60\% in 2016, as shown in Figure \ref{fig:ASASN}. Similarly, the semi-amplitude of the best-fit sinusoid to the TESS data varies from $3.726 \pm 0.007$\% in Sector 1 to 4.261 $\pm 0.008$\% and $4.553 \pm 0.008$\% in Sectors 27 and 28 respectively. These changes in the amplitude of the star's variability offer evidence for starspot evolution over the ASAS-SN and \TESS baselines.

\begin{figure*}[ht]
  \begin{center}
    \includegraphics[width=0.6 \textwidth, trim={0cm 0.0cm 0cm 0cm}, clip=true]{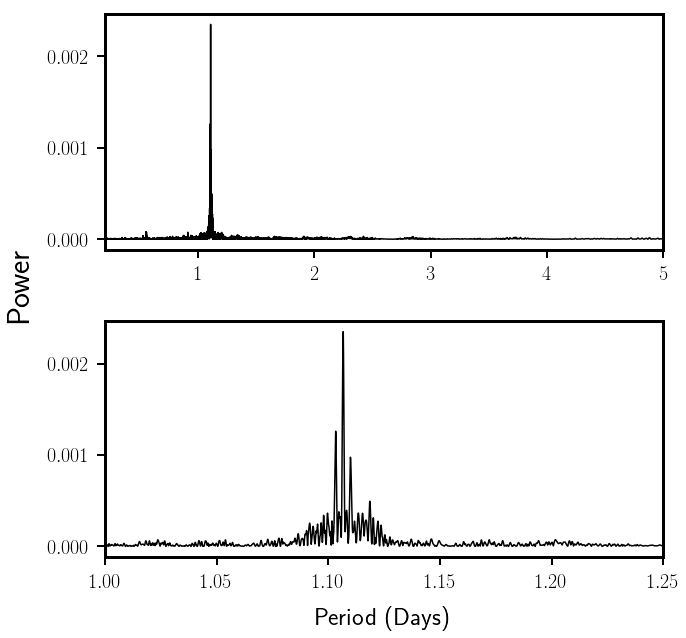}
   \end{center}
  \caption{Power spectrum of \thisstar's rotational signal, produced from \tess\ and ASAS-SN observations. We find that the star has a rotational period of 1.1066 $\pm$ 0.0003 days. This is consistent with the orbital period that we have established for the dip, suggesting that the orbiting material is nearly co-rotating.}\label{fig:periodogram}
\end{figure*}

For a more precise estimate of the period of this rotational signal, we used the \lightkurve\ \citep{lightkurve} package's Lomb-Scargle periodogram  to find the maximum of the highest peak of the power spectrum of our star's rotational signal. To produce this periodogram we computed two separate power spectra, one for the ASAS-SN data and one with the \TESS\ data to minimize aliases produced by their different cadences and observing strategies. After interpolating onto a linear grid, the product of these two power arrays produced the final periodogram, which is shown in Figure \ref{fig:periodogram}. Calculating the corresponding uncertainty using the Full Width at Half Maximum (FWHM) of this peak of the power spectrum, we found our target star's rotational period to be 1.1066 $\pm$ 0.0003 days, which is consistent with the 1.1071 $\pm$ 0.0036 day stellar rotational period that we inferred from the GP. This value is also consistent with the dip's 1.1065 $\pm$ 0.0037 day orbital period.
 
\subsection{Veloce-Rosso}\label{sec:veloce}

We observed \thisstar\ with the Veloce-Rosso spectrograph \citep{Gilbert18} at the 3.9-meter Anglo-Australian Telescope of the Siding Spring Observatory over six nights between 2020 October 27 and 2020 November 09. Veloce has a resolution of $R \approx 80,000$ and, at present, obtains data over the wavelength range 580-930 nm.

Because M dwarfs are dim and red stars, the H$\alpha$ line is the most practical proxy for magnetic activity in their convective exteriors \citep{Bell_2012}. Notably, stellar flares lead to heightened emission in the Balmer lines and other chromospheric spectral lines \citep{Kowalski_2013}, and H$\alpha$ emission has also been associated with M dwarf activity-rotation relations \citep{Newton_2017, Reiners_2012}. Accordingly, our analysis focuses on H$\alpha$, with the interpretation further discussed in Section \ref{sec:slingshot}.

Three of our exposures were taken as 20-minute observations; all other observations were 30 minutes in duration. We used the standard Veloce observing setup from its planet search program, similar to that of \citet{Bouma20}.  The airmass of these observations ranged from 1.18 to 1.49, while seeing ranged from 1\farcs7 to 2\farcs8. In total, we obtained 29 spectra. At wavelengths near H$\alpha$, the SNR of these spectra ranges from 10 to 27 per pixel with a mean SNR of 20. 

We reduced these data to extract the spectral order containing the H$\alpha$ line. After removal of the bias level and flat field, we performed a box extraction over a region 49 pixels wide to account for the 19 target fibers corresponding to a 2\farcs5 diameter region of the sky centered on our target. Veloce also obtains five sky spectra through offset fibers observed simultaneously, although we note that, at these wavelengths, the sky emission is negligible compared to the brightness of our target star. We infer a wavelength solution from observations of a Thorium-Xenon lamp  with the same instrumental setup. 

\begin{figure}[ht]
  \begin{center}
    \includegraphics[width=0.5\textwidth, trim={0cm 0.0cm 0cm 0cm}, clip=true]{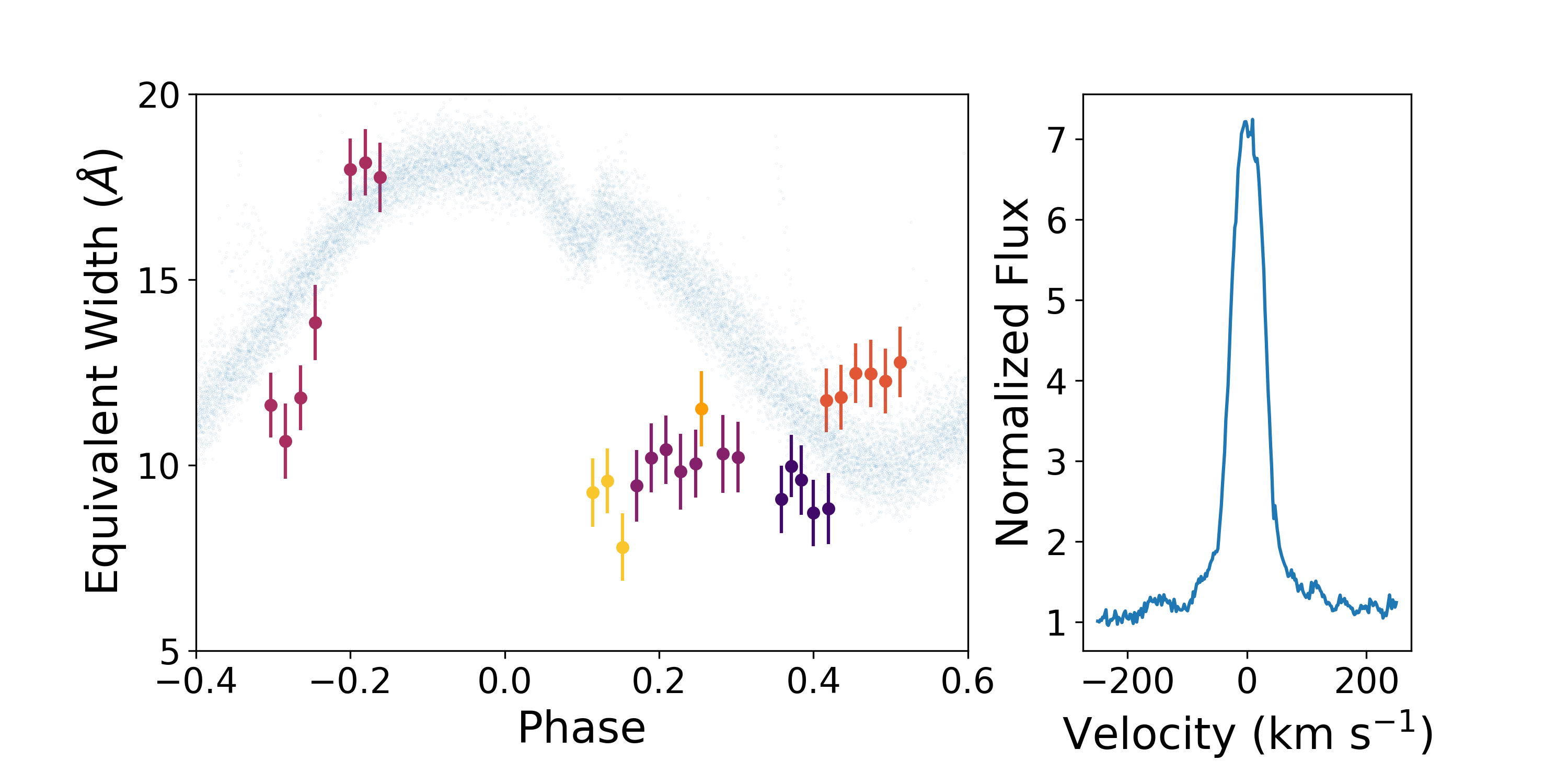}
   \end{center}
  \caption{Veloce-Rosso data. Left Panel: The H$\alpha$ line's equivalent width, with each color corresponding to a different day of observations, is plotted over the phased \tess\ data. Right Panel: The H$\alpha$ signal, co-added across all observations.}
    \label{fig:spectra2}
\end{figure}

A significant $H\alpha$ emission signal, shown in Figure \ref{fig:spectra2}, is present in all 29 spectra obtained with Veloce. Equivalent widths vary over the range $W_\alpha = -7.8$ to $-13.8$~\AA, with the exception of three sequential spectra with $W_\alpha = -17.7$ to $-18.2$~\AA. These values are consistent with the $W_{\alpha} = -9.40$~\AA\ measured by \citet{Kraus_2014} for this star, and broadly fit with the range of equivalent widths measured by \citet{Kraus_2014} for young M dwarfs in the Tuc-Hor association and by \citet{Scholz07} for M dwarfs in the slightly younger $\beta$ Pictoris moving group.

We fit a velocity-broadened template to Fe \small{I} lines near 840 nm to infer the projected rotational velocity of \thisstar, finding \vsini\ = 11.9 $\pm$ 0.4 km s$^{-1}$. This value is consistent with the 11.9 $\pm$ 0.9 km s$^{-1}$ velocity identified by \citet{Kraus_2014}. Combining this line width with the stellar rotational period inferred in Section \ref{sec:asas-sn}, we derive an angle $i_{\rm \ rot}$ between \thisstar's spin axis and \TESS's line of sight of 37.5 $\pm$ 2.0$^\circ$. We discuss the implications of this result in the context of the centrifugal breakout model in Section \ref{sec:magnetospheric}.
 
We also investigate the RV stability of these spectral features to place a limit on the mass of orbiting objects. 
We measure the centroid of individual Fe \small{I} lines across all epochs to infer a stellar radial velocity, and fit the resultant RV signal to a sinusoidal signal with period equal to that of the dips. This period is nearly identical to the stellar rotational period. Rotationally-driven modulation of $\sim100$ m s$^{-1}$ has been observed in other members of Tuc-Hor \citep[e.g.][]{Montet19}, so while a detection would not unambiguously suggest a massive companion, we can place an upper limit on the potential mass of orbiting objects. Here, we can rule out signals with a 1.1-day period and a Doppler semiamplitude $K > 730$ m s$^{-1}$ at 95\% confidence. This constraint implies that if the observed dips are caused by a transiting object, its mass must be no larger than 2.1 \mjup.

\subsection{LCO}

\begin{figure}[ht]
  \begin{center}
    \includegraphics[width=0.45\textwidth, trim={0cm 0.0cm 0cm 0cm}, clip=true]{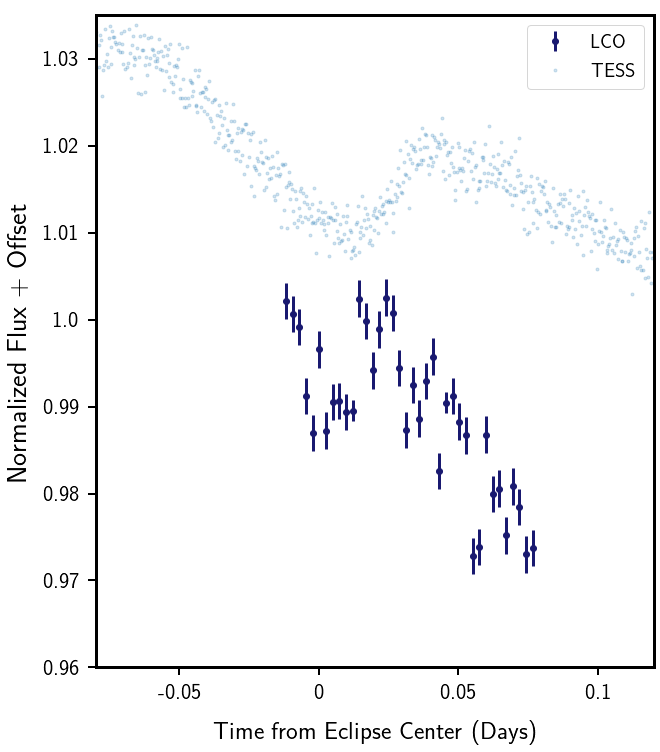}
   \end{center}
  \caption{LCO follow-up photometry is shown in dark blue with Sector 27 \tess\ data from over 100 days earlier in light blue, providing evidence for the return of a dip over the intervening three months. An offset has been introduced between the two data sets' flux values for ease of visualization. We interpret the different slope of the out-of-transit signal in the \TESS\ and LCO data as reflecting observations in different bandpasses.}
    \label{fig:LCO}
\end{figure}

We obtained follow-up time-series photometric observations of
\thisstar\ using the 1~m telescopes in the Las Cumbres
Observatory network \citep[LCO;][]{brown_cumbres_2013}. The Sinistro cameras mounted on the LCO-1m telescopes have a
$26\arcmin\times 26\arcmin$ field of view, and an unbinned pixel scale
of $0\farcs39$\,pix$^{-1}$.  

The observations were conducted on the night of UT 2020 November 13 from
the SAAO node, with mild defocusing.  We opted to use the SDSS
g$^{\prime}$ filter\footnote{\lcourl},
based on the expectation that the dips would be chromatic, with the
largest depths in the bluest bandpasses \citep[e.g. ][]{onitsuka_multicolor_2017,Tanimoto_2020,gunther_complex_2020}.
Ingress and egress were predicted for 18:54 and 20:24 (UT) on the
night of based on the \tess\ data.
% (In BJD-2400000 that's 59167.288 and 59167.350.)
The observations began at astronomical dusk (UT 18:46), continuing until UT 20:53. The
full-width at half-maximum of the image point spread function varied
between 13.6\,pix ($5\farcs2$) and 17.3\,pix ($6\farcs7$), with a
median value of 15.1\,pix ($5\farcs9$). 

We reduced the images to aperture photometry light curves using the
{\sc FITSH} package \citep{pal_2012}, with the Gaia DR2 catalog used
as a reference for determining the astrometric plate solution of each
image. We used three concentric apertures for photometry with radii of
15\,pix ($5\farcs8$), 20\,pix ($7\farcs8$) and 25\,pix ($9\farcs7$).
The background flux was estimated in an annulus about each aperture
with an inner radius of 51\,pix ($19\farcs9$) and width of 20\,pix
($7\farcs8$). For each aperture, the light curve of the target star
had an ensemble correction applied using the other sources in the
field as comparison stars. Auxilliary parameters, such as the image
position, PSF shape, background, and background deviation were also
measured, but, given the uncertainty regarding the expected form for
the astrophysical signal that would be present in the observations,
we did not detrend against these parameters.

The results are shown in Figure~\ref{fig:LCO}.  Here, a $\sim$~1 hour dip of
depth $\approx$15 mmag appears at the beginning of the
observing sequence. Fitting the dip to a transit model with the same shape as the Sector 27 events but with the depth allowed to vary as a free parameter, we measure a depth of $1.4 \pm 0.4\%$ in this filter. This event may not be associated with the Sector 27 event given the Sector 28 nondetection. Moreover, since the opacity of the transiting material is likely to have a wavelength dependence, the different  bandpasses of the two telescopes can produce apparent depth variations. 

With those caveats in mind, we also considered a standard, symmetric transit model following the formalism of \citet{Mandel02}, which produces a measured eclipse depth of $1.3 \pm 0.4\%$. Therefore, either set of assumptions produce a $3.5\sigma$ detection of a transit-like event on this night.
The detection of this event is significant
because the dip was not visible in \tess\ Sector 28, the last month of data collected prior to these LCO observations. 

\section{Analysis}\label{sec:analysis}

\subsection{Validating the Signal}

In the \tess\ bandpass, \thisstar\ is 7.3 magnitudes brighter than the closest star (28\farcs92; slightly larger than one \tess\ pixel) and 4 magnitudes brighter than any star within one arcminute. As four magnitudes corresponds to a flux ratio of 40, the rotational signal observed in both \tess\ and ASAS-SN data can only be attributed to \thisstar. If this signal were from a background star, it would require at least a 60\% obscuration of that star to produce the diluted 1.5\% dips observed in \tess\ data, and the periodic signals on two unassociated stars would need to have the same period to within 1-2 minutes, further underscoring the unlikely coincidence needed. Moreover, our LCO follow-up photometry indicates that the dips must be localized to the target star to within $\sim2''$, well within the distance of any resolved sources in the Gaia EDR3 catalog \citep{gaiacollaboration_edr3_2020}, providing additional evidence that the signal indeed belongs to our target star. 

%2.51^4 = 39.7
%0.015/(1/40)=0.60

\subsection{Eclipse Profiles}\label{sec: profiles}

\begin{figure}[ht]
  \begin{center}
    \includegraphics[width=0.5\textwidth, trim={0cm 0.0cm 0cm 0cm}, clip=true]{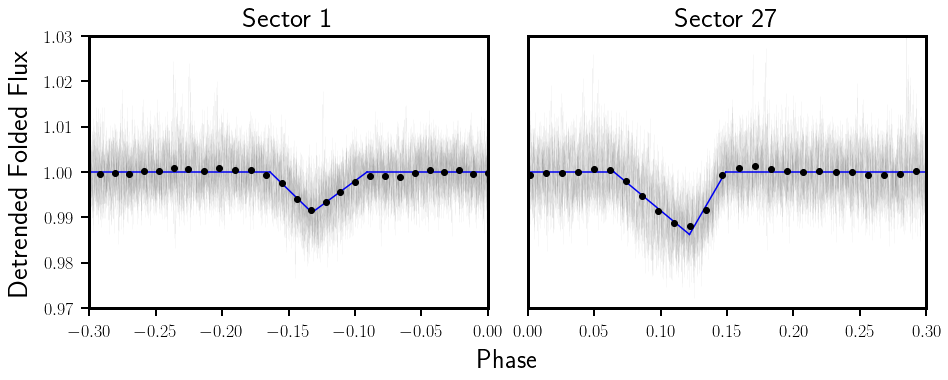}
   \end{center}
  \caption{Asymmetry of \thisstar's dips in Sectors 1 and 27. A section of the full detrended, folded \TESS\ data, with error bars, is shown in gray. A binned version of this same data is shown in black. Our best-fit triangular transit model is superimposed in blue. Note the asymmetry of the transit, with the slower half of the event occurring on the side closest to the maximum brightness of the rotational modulation in both sectors.}
    \label{fig:asymmetry}
\end{figure}

\thisstar's dips are visibly asymmetric, with a distinctively triangular shape (see Figure \ref{fig:asymmetry}). To see the asymmetry more quantitatively, we individually performed a linear regression on the ingress and egress of the folded, detrended dips in both Sector 1 and 27. We measured a slope for ingress and egress in Sector 1 of $-1.18\pm0.06$\% per hour and $0.90\pm0.06$\% per hour, respectively. In Sector 27, the measured slopes are $-1.008\pm0.025$\% per hour and $2.11\pm0.07$\% per hour, respectively. With both ingresses inconsistent with the absolute value of the slope of their respective egresses, there is significant evidence for dip asymmetry in the \tess\ data. 

% Left side of Sector 1: slope -0.283411 \pm 0.01503306, reduced chi-squared 1.83955492
% Right side of Sector 1: slope 0.21587 \pm 0.0147719, reduced chi-squared 1.90282

% Left side of Sector 27: slope -0.242522 \pm 0.00589121, reduced chi-squared 2.23809
% Right side of Sector 27: slope 0.506572 \pm 0.0165881, reduced chi-squared 1.60254

It is especially interesting to compare these results with the eclipse features of \Rik\ and the transient flux-dip stars. Like \thisstar, those stars exhibit asymmetric and triangular transits \citep{David_2017, Stauffer_2017}. In particular, this triangular shape may be taken to suggest that the corotating material has a total extent roughly comparable to the size of their host star. Such a conclusion also matches \citet{David_2017}'s estimates for \Rik, which are based on the observed transit duration. 

Interestingly, even with the phase change that occurred from Sector 1 to Sector 27, our dip is in both cases asymmetric in a similar way, with the shallower slope pointed to the peak of the starspot signal. An examination of light curves from other transient flux-dip stars indicates that they also seem to share this feature, hinting that there could be a physical mechanism---potentially related to the corotating plasma tracking magnetic activity---behind this feature. Future work will be needed to understand this in more detail. 

\subsection{Changes in Eclipse Parameters}\label{sec:dip_params}

The light curve shown in Figure \ref{fig:stacked_sectors} indicates that, apart from the sudden disappearance of the dip between Sector 27 and 28 (visible in the center panel), the variation in the depth and duration of \thisstar's dips is systematic and gradual. As Figure \ref{fig:stacked_sectors} shows, we also see the signal shift in phase between Sector 1 and Sector 27, indicating either that the orbital period does not perfectly coincide with \thisstar’s rotational period or that the two signals have different origins. 

We now analyze changes in the equivalent duration of the dips present around \thisstar. Following the methodology described in \citet{Hunt_Walker_2012} for the analysis of stellar flares, but with opposite sign conventions, the equivalent duration  $t_{\rm equiv}$ is defined by

\begin{equation}
t_{\rm equiv} = \int_{\rm t} \left(1 - F_{\rm {norm}} \right) dt,
\end{equation}
where $F_{\rm {norm}}$ is the normalized flux and the dips are integrated over all times $t$ across their duration. This definition mirrors the equivalent width in spectroscopy, with an integral over time instead of wavelength. Conceptually, the equivalent duration expresses the amount of time that the star's normalized flux would stay at zero in order for the same time-integrated amount of light to be blocked as in the actual event that we observe. Thus, the equivalent duration is a proxy for the amount of transiting material, and it is an appropriate choice for our purposes, since---as Figure \ref{fig:stacked_sectors} suggests---the duration, depth, and morphology of \thisstar's dips change simultaneously.

\begin{figure}[ht]
  \begin{center}
    \includegraphics[width=\linewidth]{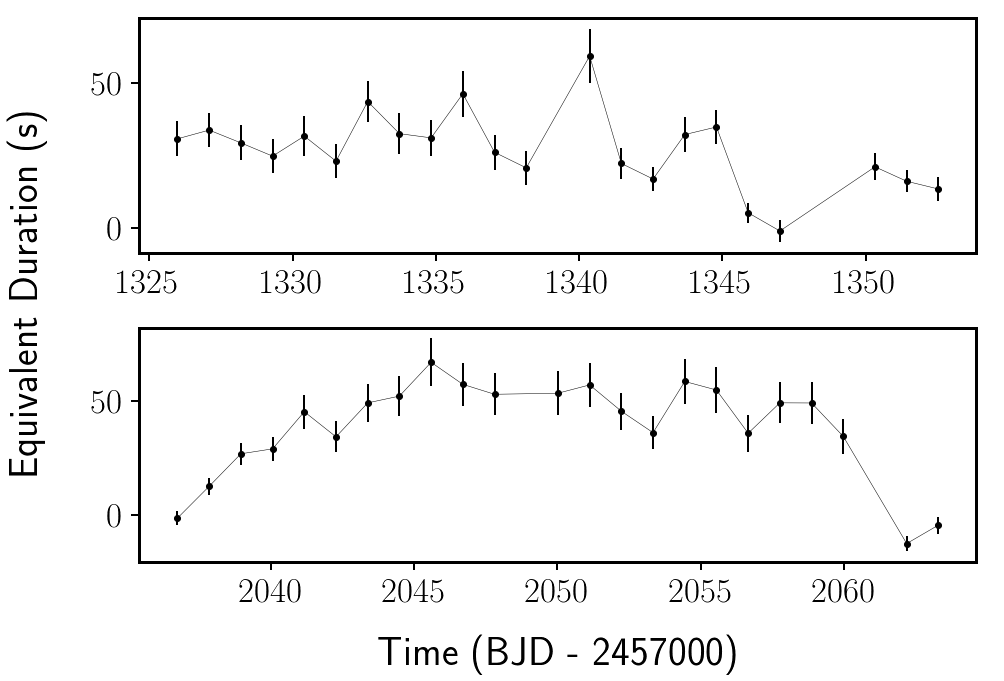}
   \end{center}
  \caption{Changes in equivalent duration, the time-integrated observed flux decrement and a proxy for the amount of material transiting the star, during \TESS\ Sectors 1 and 27. The first two orbits of Sector 28 are shown along with the Sector 27 data. Note the equivalent duration gradually decreases in Sector 1 and increases in Sector 27, followed by the sudden disappearance of any detectable dip between Sector 27 and 28.}
  \label{fig:equivdur}
\end{figure}

To calculate the equivalent duration, we integrated across the dips in our detrended data from Sectors 1 and 27, using the median of the baseline immediately surrounding each dip for improved normalization. Results are shown in Figure \ref{fig:equivdur} and indicate that variations in the equivalent duration are not purely stochastic, but rather very systematically on $\sim$20-day timescales. Moreover, as the center panel of Figure \ref{fig:stacked_sectors} indicates, we also see potential evidence of drift in the transiting material's orbital period prior to the disappearance of the dip. However, as discussed in Section \ref{sec:asynchronicity}, dip duration and morphology changes are difficult to separate out from any possible drift, so we cannot be certain of this conclusion.

\subsection{Flare Energies and Rate}\label{sec:flares}
   
\begin{figure}[ht]
\begin{center}
\includegraphics[width=0.46\textwidth,trim={0.25cm 0 0 0}]{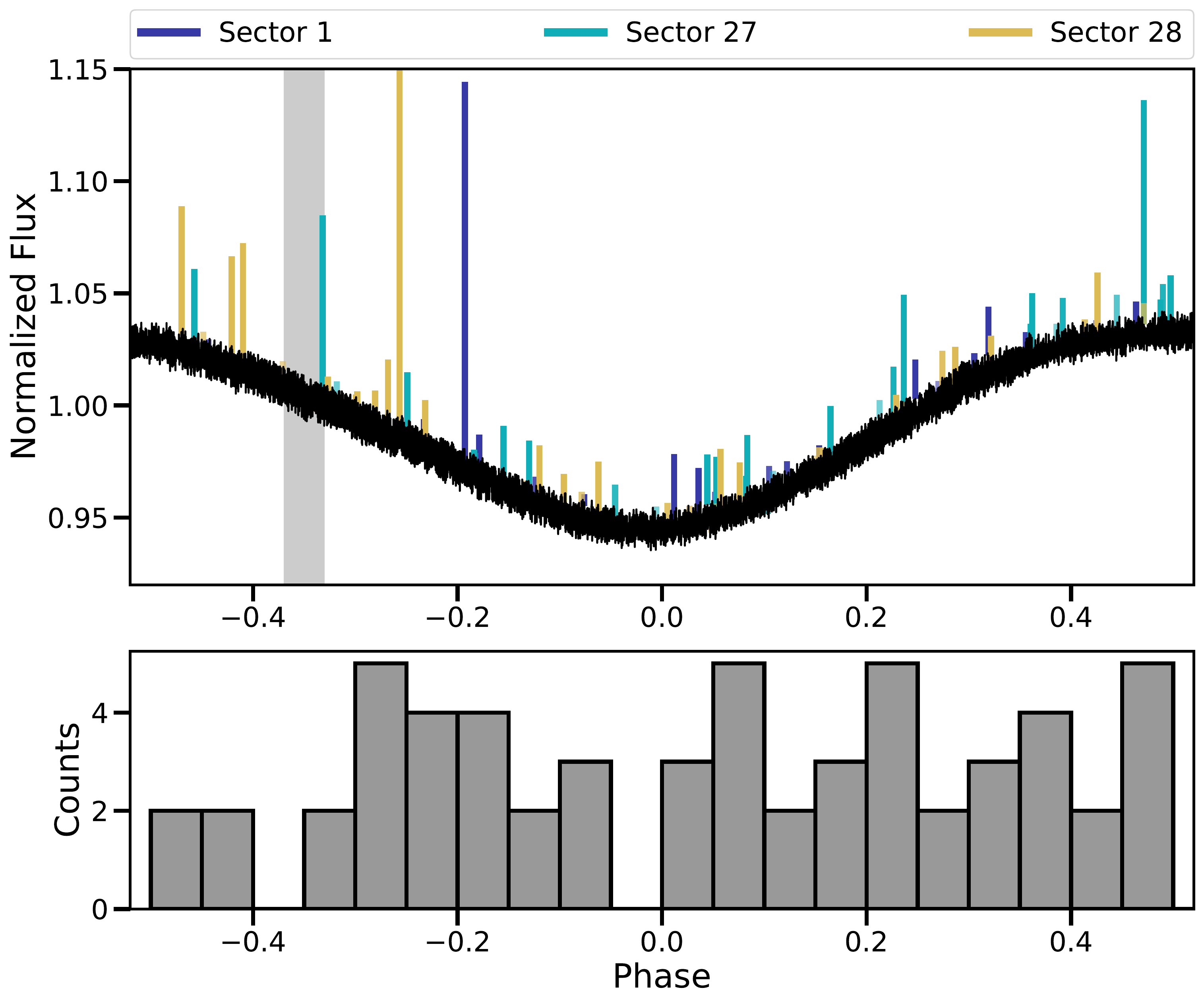}
\caption{Results of the \texttt{stella} CNNs run on the three light curves of \thisstar. Top Panel: Distribution of flares over a phase folded light curve for \thisstar. Flares are scaled by the amplitude of the flare and are colored by the \TESS\ sector they were observed in. The gray shaded region highlights the phase at which the dip is located.  The upper limit of the y axis is held at a normalized flux of 1.15, so the largest flare in Sector 28, as shown in Figure \ref{fig:bigflare}, is not visible in its entirety. Bottom Panel: A histogram representation of the number of flares identified at each rotational phase. Flares are binned in $\Delta$\,Phase = 0.05. There is no evidence for a correlation between flare rates and rotational phase.}\label{fig:flares_phased}
\end{center}
\end{figure}

To identify flares present in the light curve of \thisstar, we applied the convolutional neural network (CNN) \texttt{stella}, developed by \cite{feinstein20} and trained on flares identified by \cite{guenther19_flares}. The \texttt{stella} CNNs allow for flare detection without removing underlying rotational modulation and assigns probabilities to each identified flare being true. Following the methods of \cite{feinstein20}, we run and average the probability outputs of 10 CNN models. We included any flare with a \texttt{stella}-averaged probability $> 0.9$, which indicates the CNN models estimate a 90\% confidence this is a true flare event. The result of this analysis is presented in Figure~\ref{fig:flares_phased}.

We identified 58 flares across all three \tess\ sectors, visible in Figures \ref{fig:stacked_sectors} and \ref{fig:flares_phased}. The flare rate was calculated by weighting each flare by the probability assigned by \texttt{stella}, with an error-bar assigned by assuming that flares follow a Poisson distribution. This yields an average flare rate of $1.35 \pm 0.14$ flares per day across Sectors 1, 27, and 28. Recent work from \cite{Howard_2020} suggests that even 20,000 K flares may not be uncommon around low-mass stars, but we do not have any temperature information for individual flares given the single \TESS\ bandpass. Here we assume a flare temperature of 9000~K, as is typical of flares around the Sun \citep{Kretzschmar_2011}. With this assumption, and taking \teff\ for \thisstar\ to be 3100~K, we find that the flares in the sample span an energy range of $1.6 \times 10^{32}$ to $1.3 \times 10^{34}$\,erg, with a median energy of $9.4 \times 10^{32}$\,erg.

\begin{figure}[ht]
  \begin{center}
    \includegraphics[width=0.42\textwidth, trim={0cm 0.0cm 0cm 0cm}, clip=true]{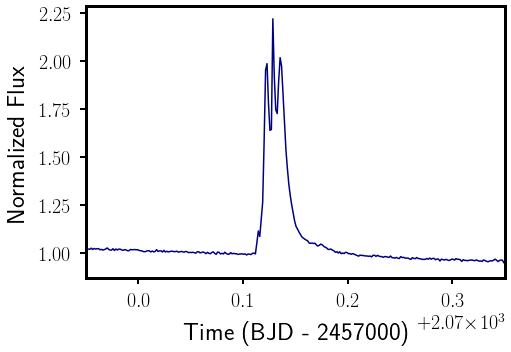}
   \end{center}
  \caption{The $\approx$120\% triple flare in the \TESS\ bandpass that appears in Sector 28, days after the disappearance of the dip. This may be interpreted as additional evidence for increased magnetic activity in the time surrounding the potential centrifugal breakout event.}
    \label{fig:bigflare}
\end{figure}

A triple-peaked flare event with an energy of $3.9 \times 10^{35}$\,erg and a measured equivalent duration of 1900.16\,s was identified at $t = 2070.128$\ BTJD (Figure~\ref{fig:bigflare}). A different flare was identified 16 minutes from the phase of the eclipse time projected from Sector 27, but there is no significant relation between the flare occurrence rate and phase overall, as shown in the bottom panel of Figure \ref{fig:flares_phased}. Although geometric considerations can lead flares to be unobserved, we note that the observed flare rate is lower in Sector 1 ($0.81 \pm 0.17$ flares per day) than in Sector 27 ($1.45 \pm 0.25$ flares per day) and 28 ($2.00 \pm 0.32$ flares per day). This quiescent period coincides with a period of relatively shallow dips, as further discussed in Section \ref{sec:mass_loss}.

\section{Potential Origins of the Eclipses}\label{sec:dips}

Here, we discuss possible origins of \thisstar's dips. We consider in turn the possibility that the dimming events are caused by  a disintegrating or sublimating planet, a precessing planet, a planet transiting an active stellar surface, an eclipsing black hole-M dwarf binary, secondary eclipses of slingshot prominences, or transiting magnetospheric clouds, determining that mangetospheric clouds are the most plausible cause of the dips.

\subsection{A Planetary Origin}

\subsubsection{A Disintegrating or Sublimating Planet}

A distinguishing feature of \thisstar's light curve is the variation in its eclipse characteristics over a few days. Beyond the sudden disappearance of the dip over the course of one rotational period between Sectors 27 and 28, there are changes in the depth, duration, and morphology of the dips throughout Sectors 1 and 27 (See Figures \ref{fig:stacked_sectors} and \ref{fig:equivdur}).

There is precedent for changes in transit parameters in planetary systems. For example, variable mass loss by disintegrating planets can produce changes in the transit depth and duration  \citep[e.g.][]{Sanchis_Ojeda_2015, Vanderburg_2015}. However, the known disintegrating planets have somewhat shorter periods, ranging from 4.5 to 22 hours \citep{Lieshout_2018, Vanderburg_2015}, more extreme and stochastic depth variations \citep{Rappaport_2012} and a pre- and/or post-ingress bump that may be ascribed to the scattering of light \citep{Rappaport_2012, Sanchis_Ojeda_2015}. 

For this system orbiting a cool M dwarf, we conservatively estimate the Roche limit by assuming a low planet density, 0.5 $\mathrm{g}\ \mathrm{cm}^{-3}$. The Roche limit in this configuration is 0.0035 AU, compared to a co-rotation radius of 0.016 AU. Moreover, the gradual change in equivalent duration that we see in Figure \ref{fig:equivdur} does not match the expectation for purely stochastic variations in the geometric arrangement of transiting dust. Based on these arguments, the dips in the light curve of \thisstar\ are inconsistent with the presence of a disintegrating planet. 

% The corotation radius is calculated in Section 5.1 

% For the Roche limit d: 

% $$d = R_* \left(\frac{2 \rho_*}{\rho_m}\right)^{1/3}$$ where $R_* = 2.975 \times 10^8 m $ is the radius of \thisstar, which I take to have density $\rho_* = 1.4 g\ cm^{-3}$, as for the Sun.

We next consider the possibility that the dips are instead caused by a sublimating planet. As in Section \ref{sec:stellar_params}, we estimate the temperature of material at the Kepler co-rotation radius to be $T_{\rm K}= 806 \pm 41$ K. The transiting material is thus below the sublimation temperatures of olivine, pyroxene, and carbon, although likely above the sublimation temperature of iron \citep{Kobayashi_2011}. This implies that the signal is likely not from the sublimation of a planet with an Earth-like composition, although a strictly iron core could in principle be sublimating. However, this scenario would not cleanly explain the co-rotation between the orbiting material and the stellar rotation or the phase change from Sector 1 to 27. Moreover, from Sector 28 data, we can rule out a dip deeper than 400 ppm, and therefore the presence of a planet bigger than 0.9 \rearth, with 95\% confidence. Hence, any material that we observe in Sector 27 is dominated by a component other than an opaque transiting planet, if one exists in the system.

% \
% $$T_{corot}=0.34\ T_{eff} \left(\frac{P}{1\ day}\right)^{-\frac{1}{3}}\left(\frac{R_*}{R_\odot}\right)^{\frac{1}{2}}\left(\frac{M_*}{M_\odot}\right)^{-\frac{1}{6}}$$ 

% $$=0.34\ 3100 K \left(\frac{1.1066175\ days}{1\ day}\right)^{-\frac{1}{3}}\left(\frac{0.427784\ R_\odot}{R_\odot}\right)^{\frac{1}{2}}$$
% $$\left(\frac{0.422456\ M_\odot}{M_\odot}\right)^{-\frac{1}{6}}$$ 

% $$=(806 \pm 41)\ K$$

\subsubsection{A Precessing Planet}

A precessing planet is a planet that is being torqued in and out of our line of sight by a massive object; this is another possible cause of the depth-varying dips that we observe, particularly since the dip disappears in \tess\ Sector 28. However, transits of precessing planets vary over timescales around two orders of magnitude larger than what is seen for \thisstar. For example, \PTFO, if explained as a transiting planet, is thought to have a precession period of 293 days or 581 days \citep{Barnes_2013}. Meanwhile, K2-146, a mid-M dwarf with two transiting planets, shows extreme transit-timing variations and has an estimated nodal procession period of 106 years \citep{Hamann_2019}. Since the dip disappears over a $\sim$1-day timescale, precession alone cannot explain the sudden disappearance of the dip between Sectors 27 and 28. Additionally, a precessing planet does not neatly explain the match between the orbital and rotational periods or the phase change of the signal from Sector 1 to Sector 27.

\subsubsection{A Planet Transiting an Active Stellar Surface}

Another possibility is that the dips originate from a planet transiting an active latitude on \thisstar. Under this scenario, the planet's transit chord would block 
a region on the stellar surface that has variable flux. For example, Sector 1's gradual decrease in the dip depth over time could correspond to a planet that is transiting across a band of starspots, where that band is gradually growing to cover a larger fraction of the star's surface over time; Sector 27's increase in dip depth would then correspond to a decay of a similar band. 

For this situation to explain the transit duration variations that we observe in Sector 1 and 27, there would need to be a sharp boundary between the active region and the quiet region on the star. Otherwise, we would observe a dip that continues to have a consistent duration (due to the fixed size of the planet), despite varying depth during and between transits. Moreover, the gradual changes in dip equivalent duration that we observe (Figure \ref{fig:equivdur}) over the course of Sectors 1 and 27 imply that the active region must be moving slowly across the stellar surface. Finally, since we do not observe significant changes in the spot signal over the sector, the active band would likely have to be rotationally symmetric, and therefore an active latitude. 

However, the sudden disappearance of the dip at the Sector 27 to 28 boundary implies that this hypothetical active region would have to grow from a near-minimum state to its maximum state within a $\sim$1-day timescale---at odds with the slow growth required to explain the changing dip parameters in Sector 1 and 27. Moreover, given that we do not see any change in the rotational signal, this scenario would require that the rapid growth of the active latitude occurs uniformly, across all longitudes of the star. Long-term ASAS-SN data does not give any evidence for step-function magnitude changes at this level, so we can disfavor this model.

\subsection{A Compact Companion}\label{sec:blackhole}

In principle, the stable and strong sinusoidal signal that we have, until now, attributed to \thisstar's starspots could instead be explained by relativistic Doppler beaming. In this scenario, light emitted isotropically in a moving object's reference frame becomes concentrated when moving towards an external observer and fainter when moving away, leading to an ellipsoidal modulation in the light curve \citep[e.g.][]{Mazeh_2010, Faigler_2013}.

To first order, the amplitude of the beaming signal is proportional to $4K/c$, where $K$ is the orbital Doppler semi-amplitude and $c$ the speed of light. In the context of our system, beaming would require \thisstar\ to be a binary system with a black hole. Under this scenario, the transit-like signal would be caused by the secondary eclipse of \thisstar\ by its compact companion, and the changing dip parameters could be explained due to varying levels of accretion onto the black hole. 

We reject the compact companion hypothesis for two reasons. First, Doppler beaming would require an unchanging sinusoidal signal because the binary system would be expected to have fixed orbital parameters that would lead to a perfectly repeatable beaming effect. This is not, however, consistent with our observations. As described in Section \ref{sec:asas-sn}, the observed amplitude varies by more than a factor of two from year to year, suggestive of starspot evolution and inconsistent with Doppler beaming. 

% Year, amplitude, uncertainty on amplitude for 5 years of ASAS-SN data: 
%[(2018, 0.058, 0.013),
%  (2017, 0.0682, 0.0065),
%  (2016, 0.0745, 0.0060),
%  (2015, 0.0560, 0.0074),
%  (2014, 0.0273, 0.0077)]

% Sector, amplitude, uncertainty from same fit to TESS data: 
% Amplitude and Uncertainty from fitting a sine curve to TESS data:
% Sector 1:  0.037164 $\pm$ 6.7e-05
% Sector 28: 0.045528 $\pm$ 8.1e-05
% Sector 27: 0.042612 $\pm$ 7.7e-05

Moreover, if \thisstar\ were in a binary system, the semi-amplitude of the photometric variability in the \TESS\ data should correspond to a radial velocity signal with an amplitude on the order of 2300 km s$^{-1}$, as inferred from Equation 2 from \cite{Loeb_2003} and accounting for bandpass effects after the methods of \cite{Herrero_2014}. We do not observe such variability in the spectra. This scenario would also require the black hole to have an implausibly large mass of over 1000 M$_\odot$ \citep{Loeb_2003}. Hence, we can effectively rule out both the presence of a compact companion and this scenario for the origin of the dips.

% Veloce would have been sensitive to a Doppler signal with that amplitude; similarly, the $v\ \sin(i)$ obtained by \cite{Kraus_2014} was on the order of only 10 km s$^{-1}$.

\subsection{Secondary Eclipses of Slingshot Prominences}\label{sec:slingshot}

Slingshot prominences are cool, dense, co-rotating clumps of gas that are trapped along coronal field lines \citep{Jardine_2018}. Prominences are especially common around stars with strong magnetic fields, such as young M dwarfs \citep{Jardine_2018} like \thisstar. Importantly, the slingshot prominence model has an associated mass-balancing mechanism that is analogous to centrifugal breakout of trapped corotating material; we could potentially use such a mechanism to explain the sudden disappearance of the dip between \tess\ Sectors 27 and 28. According to the model presented in \cite{Jardine_2019}, slingshot prominences will continue to grow until the accumulated mass exceeds the limit of what the star's magnetic field can constrain. At this point, the co-rotating material will be expelled. 

In the slingshot prominence model, the dips are caused by corotaing plasma that is emitting in specific spectral lines; when this plasma passes \textit{behind} the star, there is a decrease in the flux that we observe. One prediction of this model, would be extreme variations in the shape of H$\alpha$ as this plasma rotates across the stellar surface. These variations manifest themselves as changes in the line profile via the Rossiter-McLaughlin effect, as the blue-shifted and then red-shifted hemisphere of the rotating star is obscured in turn \citep{Rossiter24, McLaughlin24}.

However, our LCO data's bandpass is bluer than 656 nm and therefore does not cover H$\alpha$. Moreover, the \tess\ bandpass is 4000 \AA\ wide \citep{Ricker_2014} so the $5$ \AA\ change in  equivalent width that Veloce observed (See Section \ref{sec:veloce}) can only account for a 0.1\% change in the flux --- an order of magnitude smaller than the 1\%\ dip that we observe. Accordingly, it seems that some other source of emissivity beyond H$\alpha$ is necessary to account for the observed depth of the dip under the slingshot prominence scenario. In analogy with \cite{David_2017}'s reasoning for \Rik, we conclude that Paschen-continuum bound-free emission could produce broadband dimmings of up to a few percent---deep enough to produce our observed dips around for \thisstar, although insufficient to explain \Rik's dimmings.

%We reject an alternate source of emissivity---hot plasma emitting as a blackbody---because otherwise we'd expect to see both a primary and secondary eclipse.

\begin{figure*}[ht]
  \begin{center}
    \includegraphics[width=0.80 \textwidth, trim={0cm 0.0cm 0cm 0cm}, clip=true]{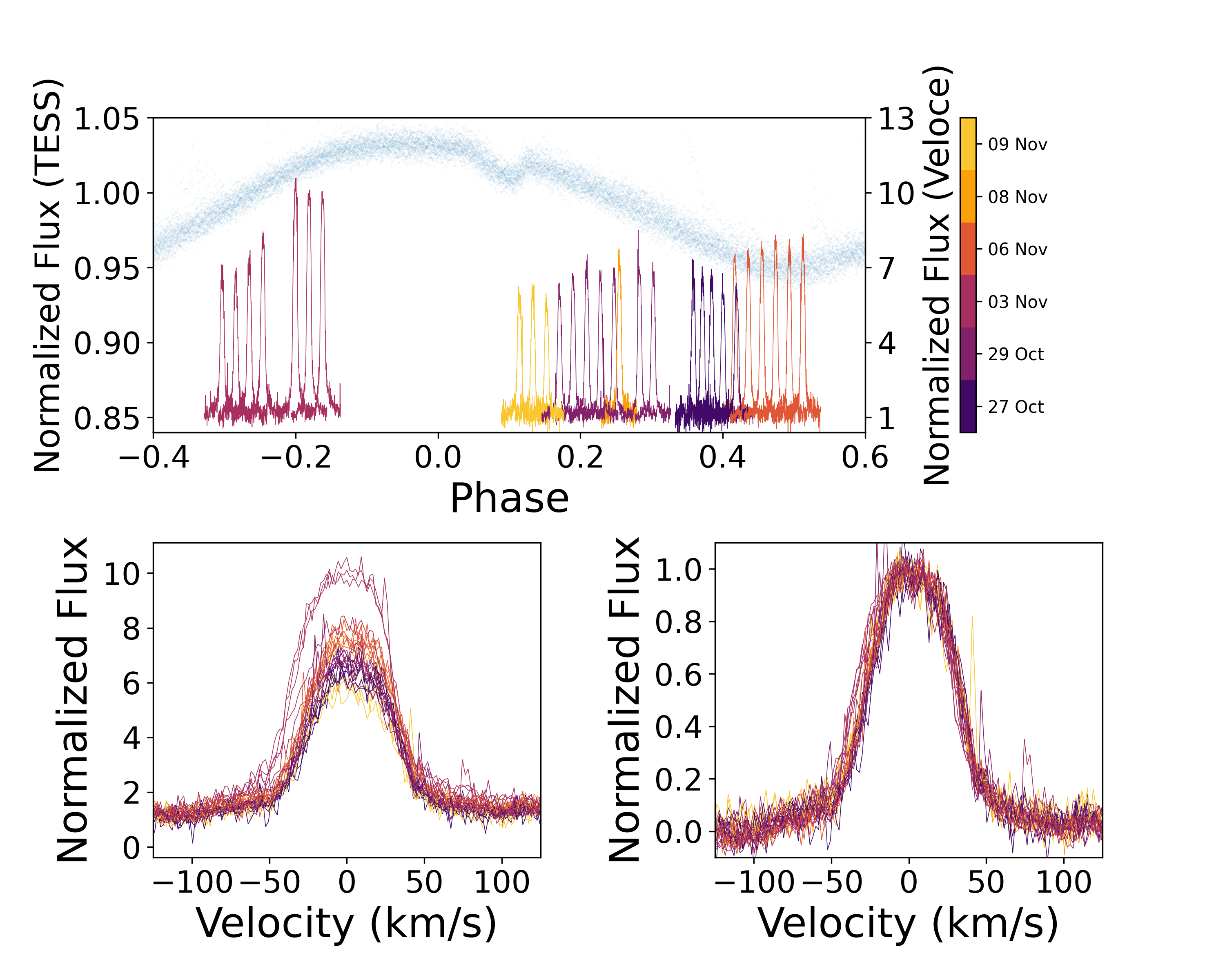}
   \end{center}
  \caption{Top Panel: Phase-folded light curve for Sector 27, with normalized H$\alpha$ spectra plotted at the phases of observations. Bottom Left Panel: The same H$\alpha$ spectra, overplotted. There are no significant shape variations over this spectral feature, leading us to disfavor the eclipsing slingshot prominence scenario. For both frames, colors correspond to the night of observation, as indicated by the colorbar on the right. The amplitude of this feature varies with starspot coverage, as is typical for young stars. There are three spectra, observed sequentially, with significantly larger amplitudes and excess emission at blue wavelengths compared to all other observations; we attribute this variability to a potential flare event on this star during these observations. Bottom Right Panel: The same as the previous subfigure, with all spectra excluding the three belonging to the flare candidate normalized to the same amplitude and equivalent width to demonstrate no significant shape variations over the remaining observations. Since the shape of the H$\alpha$ signal is very stable, the equivalent width maps consistently to the height of the spectral line, so we exclude it to avoid redundancy.}
    \label{fig:spectra}
\end{figure*}

As described in Section \ref{sec:veloce}, we observe a significant $H\alpha$ emission signal over all 29 spectra obtained with Veloce.  The observed variability, shown in Figure \ref{fig:spectra}, is largely coherent with the rotational modulation due to starspots, with a larger equivalent width observed when the starspot coverage in the visible hemisphere is higher. However, the seven observations from UT 2020 November 03 provide an exception to this trend. On this night, seven spectra were obtained: four spanning two hours and three additional spectra obtained after a 45-minute gap, spanning 90 minutes. The three spectra after the gap all exhibit an equivalent width in the H$\alpha$ line more than 20\% larger than any other feature. Moreover, during these three exposures the feature has an extended blue-shifted component with a velocity of 30-50 km/s.

This behavior is morphologically similar to spectroscopic signatures of flares \citep{Honda18, Maehara21}. In principle, it could also be emission from a co-rotating blue prominence; there is a gap of at least three days before and after these spectra were obtained. This would correspond to a shorter event timescale than those observed in Sectors 1 and 27 of \tess, but without simultaneous photometry we cannot rule out this scenario.

However, given the high flare rate for this star and the lack of line profile variations in the core of the H$\alpha$ line for any spectra obtained over this 13-night baseline, we find a stellar flare to be the most plausible explanation of the increased equivalent width and blueward asymmetry on these nights. With that caveat, there is no significant change in the shape of the H$\alpha$ feature in time, a result which is counter to predictions for a slingshot prominence \citep[e.g.][]{Zaire21}. The stability of the shape of the H$\alpha$ line, which we would expect to be mirrored in a similar stability of the Paschen lines, thus suggests a different mechanism than the slingshot prominence model.

\subsection{Transiting Magnetospheric Clouds}\label{sec:magnetospheric}

Like slingshot prominences, magnetospheric clouds consist of material originating from the star that gets trapped in the host star’s magnetosphere. However, the effects of eclipsing slingshot prominences would be primarily driven by hot plasma \citep{waugh2021slingshot}, whereas transiting magnetospheric clouds trap dust as well as ionized hydrogen, leading to additional opacity visible in broadband photometry \citep[e.g.][]{David_2017}. Since only a small amount of dust is needed to create significant optical dimmings, a magnetospheric cloud’s photometric signature can occur during a transit rather than during a secondary eclipse.

As with slingshot prominences, magnetospheric clouds would produce a signal with a period matching the rotational period. As material fills and seeps out of the magnetically-confined region of material, the depth of the eclipses can vary over time. As the material is gas and dust grains, it will have a relatively weak wavelength dependence compared to the strong spectral signature of a plasma in the slingshot prominence scenario. These predictions match the variable light curve and unchanging line profile variations observed in this data set, making this scenario our preferred explanation for the observed behavior of \thisstar. 

For \Rik, a potential younger analog of \thisstar, \citet{David_2017} also preferred the magnetospheric cloud model over the slingshot prominence model, finding that  eclipses of prominences could not explain \Rik's 20\% deep dips in the Kepler bandpass. Although the dip depth of \thisstar\ ($\leq 1.2$\%) is not large enough to conclusively rule out the slingshot prominence scenario on these grounds, the observed lack of variability in the H$\alpha$ line profile still gives us reason to prefer the magnetospheric cloud scenario over the slingshot prominence model. 

Simulations from \cite{Townsend_2008} demonstrate that a signal similar to the eclipse events that we observe around \thisstar\ can be produced with an inclination angle $i_{\rm rot} \approx 40^{\circ}$ and a magnetic obliquity --- the angle between the star's spin axis and magnetic field axis --- $\beta \approx 70-80^{\circ}$. These models are not specific to a star with \thisstar's radius and rotational period, so the simulated inclinations and obliquities may not translate exactly to predictions for \thisstar. However, they do match the observed inclination value of $i=37.5\pm 2.0$ degrees from \thisstar's radius, projected rotational velocity, and rotation period, providing confidence that relatively low inclinations and high obliquities can produce morphologically similar light curves to those collected by \tess. 

If the signal observed in this system is indeed the result of magnetospheric clouds, then we require an explanation of how sufficient material can become trapped in the magnetosphere of \thisstar\ and of how this material can dissipate over the 1-2 rotation periods between Sectors 27 and 28. These are discussed in turn below.

\section{Centrifugal Breakout}\label{sec:breakout}

\subsection{The Case for Breakout}

\begin{figure}[ht]
  \begin{center}
    \includegraphics[width=0.5\textwidth, trim={0cm 0.0cm 0cm 0cm}, clip=true]{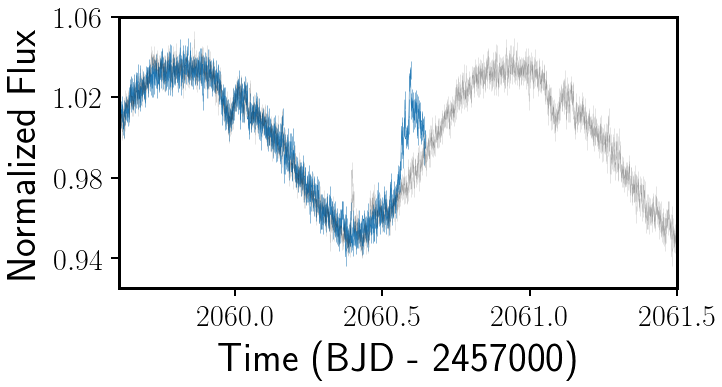}
   \end{center}
  \caption{\tess\ light curve at the end of Sector 27 (Blue), with an observed brightening at the end of the sector. For comparison, the light curve from one rotational period earlier is underlaid (Gray). Note that this brightening event has a symmetric morphology unlike the steep rise and exponential decline typical of stellar flares. We consider this feature a potential post-breakout magnetic reconnection event, highlighting that the potential analogs of \thisstar\ discussed in \cite{Stauffer_2017} also exhibited symmetric flare-like events at state transitions.}
    \label{fig:non_flare}
\end{figure}

Perhaps the most distinguishing feature of \thisstar's dips is their sudden disappearance over a $\sim$1 day timescale. Immediately prior to the disappearance of the dip, we observe a flare-like event with an unusually symmetric morphology (See Figure \ref{fig:non_flare} and Section \ref{sec:TESS}).  Moreover, in the days surrounding the dip’s disappearance, we see additional evidence for increased magnetic activity, including a $\approx$120\% triple flare some days after the potential breakout event (Figure \ref{fig:bigflare}) as well as a higher amplitude variability of the starspot signal in Sectors 27 and 28 as compared to Sector 1 (Figures \ref{fig:stacked_sectors} and \ref{fig:detrended}). In addition to the dip’s sudden disappearance, we also see evidence of variability on top of a rotational signal that is strong and stable over timescales of years. In particular, we see systematic variations in the equivalent duration of \thisstar's eclipses (See Figure \ref{fig:equivdur} and Section \ref{sec:dip_params}), a phase change of the seemingly corotating dip from Sector 1 to Sector 27, and a reappearance of the dip in our LCO data. 

All of these data are consistent with a magnetospheric cloud and centrifugal breakout model. The short timescale involved in the dip’s disappearance is plausibly explained by a sudden snapping of the star’s magnetic field lines when the mass of the corotating material exceeds \thisstar’s capacity to restrain it, whereas the alternative mass-balancing mechanisms discussed in Section \ref{sec:mass_loss} would predict longer timescales. Meanwhile, the heightened magnetic activity around the time of the dip’s disappearance would also be expected from centrifugal breakout and the unusually symmetric flare-like event that coincides with the dip's disappearance (Figure \ref{fig:non_flare}) could plausibly be a post-breakout magnetic re-connection event. 

In support of this last point, the existing theories of stellar magnetism predict that magnetic reconnection events could resemble flares \citep{Townsend_2013, Stauffer_2017}, although a link between optical flaring and reconnection events specific to centrifugally supported magnetospheres has yet to be definitively established \citep{Townsend_2013}. Suggestively, like \thisstar, some of the persistent flux-dip stars and scallop shell stars had unusually symmetric flare-like events that occurred at state transitions \citep{Stauffer_2017}. Since these state transitions and the corotating cloud origin promoted by \cite{Stauffer_2017,Stauffer_2018, Stauffer_2021} fit many of the same criteria for centrifugal breakout that \thisstar\ does, this correlation between symmetric flares and major changes in the dip properties could be taken to be an indication of the magnetic reconnection events that are expected to occur immediately after the magnetic field lines are broken during a breakout event.

Moreover, the physical mechanism of centrifugal breakout fits with the magnetospheric cloud dip origin that Section \ref{sec:dips} found to be the most probable, and \thisstar’s rotational signature, spectral type, and youth suggest that the star does, in fact, likely have a strong and stable magnetic field --- one of the prerequisites for a corotating magnetosphere.  Furthermore, the systematic variations observed in the equivalent duration of \thisstar's transit-like events (See Figure \ref{fig:equivdur} and Section \ref{sec:dip_params}) is one of the indicators of breakout that \citet{Townsend_2013} sought in their non-detection of centrifugal breakout around \sig. Finally,  a comparison with \cite{Morin_2010} suggests that the magnetic field topology of the star could plausibly have evolved over the \tess\ baseline, leading to the observed change in dip phase observed between Sectors 1 and 27. Alternatively, the phase change could be attributed to anisotropic mass loss from coronal mass ejections, as further discussed in Section \ref{sec:mass_loss}.

Regardless of the cause of the phase change, centrifugal breakout of magnetospheric clouds provides the most compelling explanation for the sudden disappearance of the observed dip. This is the only model of the possibilities considered in this work that can fully explain the observations and that matches theoretical predictions of an expected signal \citep[here,][]{Townsend_2008}. 

\subsection{Constraining the Pre-Breakout Mass}\label{sec:mass_calc}

Appendix A2 of \citet{Townsend_2005} estimates the asymptotic mass, or mass required for breakout to occur, as

\begin{equation}\label{eq:asymptotic mass}
m_\infty \approx \frac{\sqrt{\pi}\ B_*^2\ R_*^4}{6\ g_*\ r_K^2}
\end{equation}

\noindent where $B_*$ is the strength of the star's magnetic field, $R_*$ is the radius of the star, and $g_*$ is the star's surface gravity, and  $r_K$ is the Kepler corotation radius. 

In the absence of direct measurements of \thisstar's magnetic field strength, we note that observations by \cite{Shulyak_2019} would predict a magnetic field strength of around $4\pm 2$ kG for an M dwarf with \thisstar's rotational period. Accordingly, we tentatively take $B_* = 4$ kG for an order of magnitude estimate of \thisstar's magnetic field strength. Using this value together with the stellar parameters listed in Table \ref{tab:stellarpaarameters} and Equations \ref{eq:asymptotic mass} and \ref{eq:Kepler_radius}, we find the asymptotic mass to be only $\sim10^{21}$g, which is about half the mass of Saturn's satellite Janus and eight orders of magnitude below the upper limits given by the observed SED (Section \ref{sec:stellar_params}). This demonstrates that only a small amount of mass is needed to reach breakout conditions. This material may originate from the last remnants of a dissipating protoplanetary disk and could potentially be replenished by material falling off of young objects --- such as comets or forming planets.

It is also a useful exercise to compare the theoretically predicted asymptotic mass with the mass that we would expect based on the observed depth of the dip immediately prior to breakout. Here we use an approach inspired by \citet{Boyajian_2016} to estimate a lower limit for the mass of dust present immediately before the possible breakout event from the depth of the dip. We then make an order of magnitude approximation of the total minimum mass of transiting material, based on our observational lower bound on the mass of optically thick material.

First, Equation 4 from \citet{Boyajian_2016} gives 

\begin{equation}
\sigma_{\rm tot} = v_t\ h \int\tau(t)\ dt 
\end{equation}
to estimate the cross-sectional area $\sigma_{\rm tot}$ of the optically thick corotating material. Here, $v_t$ is the transverse velocity of the material, $h$ the height of the material perpendicular to its velocity, and $\tau(t)$ the optical depth as it changes over the course of a rotational period.

For $h$, we assume a spherical shape for the corotating material and use 1.2\% as the approximate depth of the last dip in Sector 27. For $v_t$, we assume uniform circular motion so that $v_t = {2 \pi r_K}/{P}$, where P, the orbital period of the material, coincides with \thisstar's rotational period and the orbital radius of the material is the Kepler corotation radius. Finally, to approximate $\int \tau(t)\ dt$ for the dip immediately prior to breakout, we use the approximation that $ \tau \approx$ ln(normalized flux). We detrend the normalized light curve as described in Section \ref{sec:TESS} and then integrate numerically over the dip. In this way we find that $\sigma_{\rm tot} = 4 \times 10^{14}$ m$^2$.

As in \citet{Boyajian_2016}, we calculate a lower limit for the mass of dust transiting the star that is given by
\begin{equation}
M_p = \frac{2}{3}\ \sigma_{\rm tot} \rho D
\end{equation}

\noindent where we take the dust grains to be made of particles with a uniform density $\rho =3$ g cm$^{-3}$ and diameter $D=1\ \mu$m. 

In this way, we would obtain an equivalent radius of $400$ m if all of the dust were gathered into a sphere with density 3\ g\ cm$^{-3}$, and we estimate a dust mass of 9 $\times$ 10$^{14}$ g. If we assume a typical 100:1 gas to dust ratio in protoplanetary disks, this gives us a minimum mass of 9~$\times$~10$^{16}$~g---about a quarter the mass of Halley's comet---for all of the transiting material.

% Hayley's comet 0.22 \times 10^{15} kg = 2.2 \times 10^{14} kg = 2.2 \times 10^{17} kg$

% $$M_p = \frac{2}{3}\ \sigma_{tot} \rho D$$ 
% $$= \frac{2}{3} 4.294 \times 10^{14}\ m^2 \times 3 \times 10^3 kg\ m^{-3} \times 10^{-6} m $$
% $$= 8.588 \times 10^{11} kg$$
% $$= 8.588 \times 10^{14} g$$

% To find its radius if it were gathered into a sphere of density $3 g/cm^3$: 

% $$M = \frac{4 \pi R^3}{3} \rho$$

% $$\frac{3M}{4\rho \pi} = R^3$$

% $$R = \left(\frac{3M}{4\rho \pi}\right)^{1/3}$$

% $$R = \left(\frac{3\times 8.588 \times 10^{11} kg}{4\ 3 \times 10^3 kg\ m^{-3}  \pi}\right)^{1/3}$$

% $$R = \left(0.6834 \times 10^8 m^3\right)^{1/3}$$

% $$R = \left(6.834 \times 10^7 m^3\right)^{1/3}$$

% $$R = 408.8 m$$

% For a first-order approximation for the total mass: 

% $0.015 \times M_{tot} = 8.588 \times 10^{14} g $

% $ M_{tot} = \frac{8.588 \times 10^{14} g}{0.015} = 573.5 \times 10^{14} g = 5.735 \times 10^{16} g $

This value is four orders of magnitude below the asymptotic mass calculated above. However, in producing the observation-based minimum mass, we have assumed that we see the entire cloud transiting, and we also do not have well-constrained estimates for the spatial distribution, diameter, composition, and density of the transiting material. Similarly, in the previous calculation we do not have a precise measurement of the magnetic field; as the mass scales as $B_\star^2$ this is a large source of uncertainty. Hence the disagreement between observation and theory suggests that either we are only seeing a small fraction of the plasma transit, the stellar magnetic field is lower than expected, or the true dust mass fraction is lower than predicted. Future work will be needed to fully explain this discrepancy.

\subsection{Asynchronicity of the Signal}\label{sec:asynchronicity}

\begin{figure*}[tb]
  \begin{center}
    \includegraphics[width=0.9\linewidth]{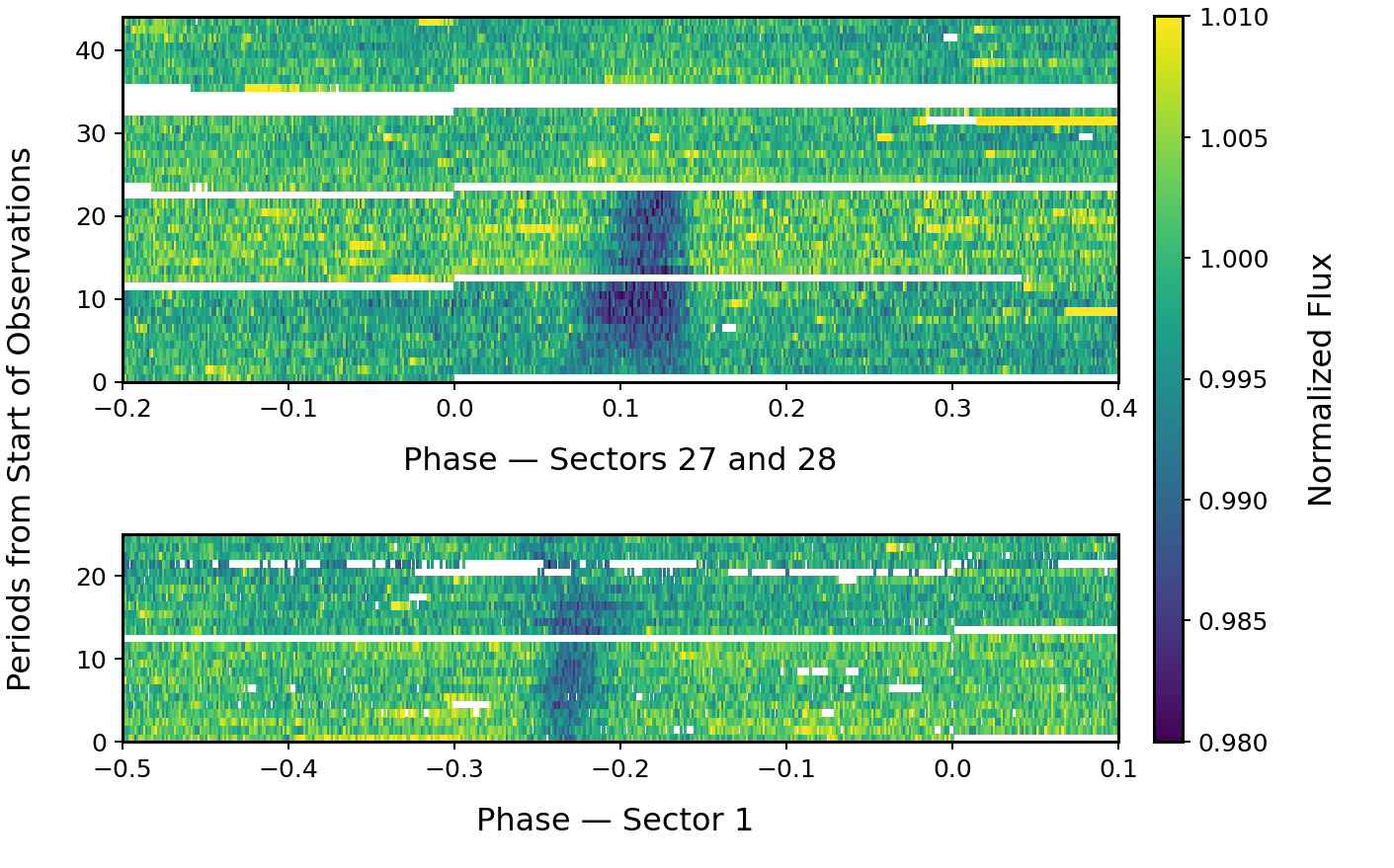}
   \end{center}
  \caption{River plot of \TESS\ data phased to the stellar spin period of 1.1066 days and with the rotational signal removed. There is an observed drift of the transit midpoint and the transit duration changes from event to event. The sudden disappearance of the dip between Sector 27 and 28, which corresponds to the potential breakout event, is also apparent.}
    \label{fig:river}
\end{figure*}

So far, we have presented the transiting material as likely to be corotating with the star. After all, the period of \thisstar\ falls within our one sigma confidence interval for the period of the dip. However, as Figure \ref{fig:river}, a river plot of the three sectors of \tess\ data, shows, the transit midpoint appears to start falling behind the predicted ephemeris as time goes on. It is not clear if this apparent drift is best explained by a change in dip duration or by a genuine change in the orbital period. In fact, even as the ingress and transit midpoint in Sector 27 begin to lag, the egress remains at a relatively constant phase throughout the sector. 

If we take this apparent drift at face value, this would correspond to 16 minutes of drift over the course of Sector 27, with a similar, though less pronounced, phenomenon occurring during Sector 1. This corresponds to a difference of less than one minute between the rotational and orbital periods. In the context of centrifugal breakout, such a drift in period could be interpreted as evidence of material gradually drifting outwards and dragging the magnetic field lines behind it until the field lines snap at the time of a breakout event. However, once again, dip duration and morphology changes are difficult to separate out from any possible drift, so we cannot be certain of this conclusion.

% The initial phase in Sector 27 is given by: 
% 20 boxes = 0.025 phases --> 8 boxes past phase 0.1, which means 8 boxes * (0.025 phases)/(20 boxes) = 0.01 phases past 0.1 --> phase 0.11

% The final phase is given by :
% 4 boxes before phase 0.125, so 4 boxes * (0.025 phases)/(20 boxes) = 0.005 phases before phase 0.125 --> phase 0.12

% The phase change is then given by 0.12 - 0.11 = 0.01 phases, and this corresponds to a drift of 0.01 phases * 1.1066 days/phase = 0.011 days = 0.011 * 24 * 60 min = 15.84 minutes 

\section{Discussion}\label{sec:discussion}

\subsection{Toward Timescale Estimates}

The evolution of \thisstar's dips appears to be governed by changes occurring on three different time-scales:

\begin{enumerate}
    \item The disappearance of the dip, which, for \thisstar, occurs on a $\sim$1-day timescale.
    \item Changes in the depth, duration, and shape of the dip, seemingly on a $\sim$10-day timescale 
    \item The post-breakout reappearance of the dip, poorly constrained by our current data, but occurring no more slowly than on $\sim$100-day timescales.
\end{enumerate}

In particular, we suspect that our current data on \thisstar\ corresponds to a minimum of three distinct dips, with one change inferred from the shift in the phase of the corotating plasma from Sector 1 to Sector 27, and another from the dip detected in the LCO data 107 days after the dip disappeared between Sectors 27 and 28. Importantly, we have no lower bound on the timescales involved in the post-breakout reappearance of the dip---and we would need such information in order to distinguish between the CME and stellar wind mass-accumulation scenarios that are presented in Section \ref{sec:CME}.

The timescales for the disappearance of the dip and for the dip depth, duration, and morphological changes appears to be broadly consistent with those for the transient flux-dip stars from \cite{Stauffer_2017}. Notably, however, for \thisstar\ we have a two-year baseline over which the dips are observed, which is important for understanding the longer-term evolution of this type of star. In particular, in these two years of data, we see a phase change, with the dip in Sector 27 offset in phase from the dip in Sector 1 by 29\% of an orbit---something observed in \PTFO\ and \thisstar, but not in any of the other known transient flux-dip stars. 

% Math for phase change: (2036.7410325-1325.98)/ 1.1066 = 642.29 -- (transit midpoint at start of sector 27 - transit midpoint at start of sector 1)/orbital period)

\subsection{Mass-Balancing Mechanisms}\label{sec:mass_loss}
Since \citet{Townsend_2013} found a lack of photometric evidence for centrifugal breakout around \sig, astronomers have been considering alternate mass-balancing mechanisms for plasma accumulating in a centrifugal magnetosphere.  In particular, \citet{Owocki_2018} presented a diffusion-drift model in which plasma escapes away from the star via diffusion and drift, and toward the star via diffusion.  \citet{Shultz_2020}, after examining early B stars with confirmed centrifugal magnetospheres and finding that their H$\alpha$ emission profiles favor centrifugal breakout over the diffusion and drift model, proposed that centrifugal breakout is the relevant mass-balancing mechanism, but that it essentially acts as a leakage mechanism. 

Meanwhile, \citet{Owocki_2020} provided an analytical framework for analyzing the observations of \citet{Shultz_2020}. In particular, \cite{Owocki_2020} proposed two possible explanations for the sudden onset of H$\alpha$ emission in early- to mid-B type stars observed by \cite{Shultz_2020}. In one explanation, the diffusion-drift model is still an important mechanism for late-B and A-type stars because the much weaker winds of these less massive stars would not allow their centrifugal magnetospheres to fill the level needed for breakout. Alternatively, stellar winds may become dominated by metal ions around stars later than mid-B and therefore lack the Hydrogen needed for H$\alpha$ emission.

We argue that \thisstar\ provides strong supporting evidence for the centrifugal breakout model, with its light curve complementing the spectroscopic and theoretical case for this model. Moreover, \thisstar\ seems to belong to the same class of stars as \PTFO\ and those studied in \citet{Stauffer_2017}, suggesting that some of the observable features of these light curves, such as the state changes and phase changes, may also be driven by the centrifugal breakout mechanism. In particular, considering the sudden disappearance of the dip around \thisstar, the centrifugal breakout candidate that we observe cannot be governed by purely continuous centrifugal leakage of the kind proposed by \citet{Shultz_2020}.

However, centrifugal breakout alone does not explain the gradual decrease in dip size observed in Sector 1, suggesting that there may still be a role for an additional mass-balancing mechanism. Although we cannot exclude a centrifugal leakage-based explanation, we consider that the corotating material will only interact with the stellar magnetic field when the dust is hot enough to stay ionized.

The dust will remain ionized while stellar flares continue to heat the corotating material. However, if  \thisstar\ enters a quiescent period, the dust will begin to undergo recombination and the apparent dip will decay. This flaring mechanism seems plausible because the work function of circumstellar dust is on the order of $\sim$5 eV \citep{tielens_2005}, so the UV radiation from flares provides sufficient energy to maintain dust in a state of ionization. This mechanism is discussed further by \cite{Osten_2013}. 

\begin{figure}[ht]
  \begin{center}
    \includegraphics[width=0.48\textwidth, trim={0cm 0.0cm 0cm 0cm}, clip=true]{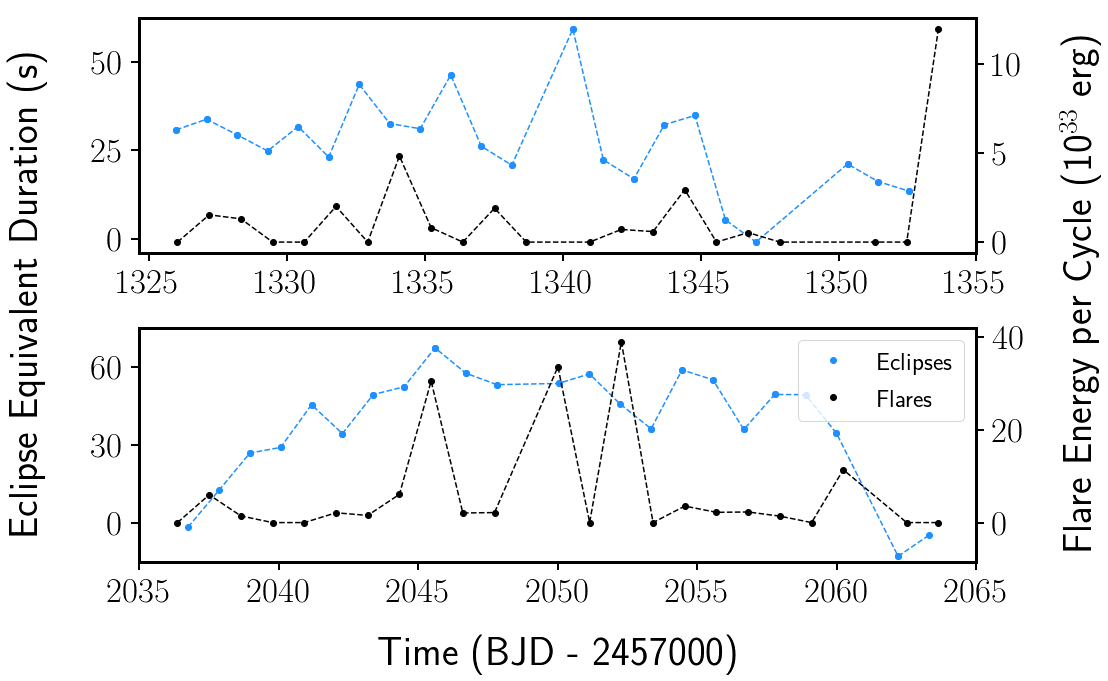}
   \end{center}
  \caption{Equivalent Durations and Flare Energies in \tess\ Sectors 1 (top panel) and 27 (bottom panel). Blue: Dip equivalent durations, as calculated in Section \ref{sec:dip_params}. Black: an estimate for the total observed flare energy released in the preceding rotational cycle, calculated using flare events identified in Section \ref{sec:flares}. Note the low flare energy in the days before the dip in Sector 1 fades away and the high flare energy surrounding the days in Sector 27 when the dip is near its deepest. Importantly, both of these observations are consistent with the flare-based reionization mechanism that we propose is responsible for the gradual fading of the dip in Sector 1.}
    \label{fig:flare_proxy}
\end{figure}

In the context of \thisstar, we note a lack of major flares in the last several days of Sector 1 (see Figures \ref{fig:flares_phased} and \ref{fig:flare_proxy}). The decay in dip size throughout the second orbit of Sector 1 is qualitatively consistent with the model proposed here. In fact, the shallowest observed eclipse in Sector 1 occurs immediately after the 10-day period with the lowest flare rate in Sector 1 and 27; similarly, the flare energy per cycle is noticeably elevated in the days surrounding the deepest dips towards the end of Sector 27 (Figure \ref{fig:flare_proxy}). Finally, there is an observed correlation (Spearman $\rho = 0.317$) between the integrated flare energy observed and the measured equivalent duration of each dip across Sectors 1 and 27, hinting at the possibility of a connection between these two observables.

\subsection{Comparison with Higher Mass Stars}\label{sec:higher_mass}

\subsubsection{Mass-Dependence of Magnetic Fields}\label{sec:Bfield}

\cite{Townsend_2005} developed the RRM model that predicted centrifugal breakout for high-mass stars like the B2 \sig, but \thisstar\ is an M dwarf. Nevertheless, the three main prerequisites for applying the RRM model, a (1) strong (kG), (2) globally organized, and (3) stable magnetic field, can be replicated with an M dwarf magnetic field.

Around magnetic hot stars, the minimum field strength for a stable large-scale magnetic field appears to occur around 300 G \citep{Auriere_2007}, with stars like \sig\ having field strengths of several kG \cite[\eg][]{oksala2015revisiting}. Similarly, observations by \cite{Shulyak_2019} would predict a magnetic field strength of $4\pm 2$ kG for a rapidly rotating M dwarf like \thisstar.

Moreover, magnetic A and B type stars often have simple magnetic fields with a globally dipole structure \citep{Briquet_2015}. Although it is common for low-mass stars to have a complex magnetic topology \cite[\eg][]{Hubrig_2019}, mid-M dwarfs often have strong, axisymmetric, dipole-dominated magnetic field configurations \citep{Kochukhov_2021}. Although future tomography observations could place tighter constraints on its spot configuration, based on its light curve, \thisstar\ appears to have a single or small number of large spot groups \citep{Basri_2018}, similar to the spot coverage that would be expected from a star with a globally organized dipolar field \citep{Shulyak_2019}.

As for stability, magnetic hot stars are thought to have ``fossil" magnetic fields that were generated early in their lifetime and that naturally exhibit almost no variability even over decades \citep{Donati_2009}. Although cool stars' deep convective zones allow for field-generating currents that maintain magnetic fields through dynamo processes and that lead to noticeable variability (for example, flares and activity cycles) on both short and long timescales \citep{Donati_2009}, M dwarfs too can have magnetic fields that are stable for over a decade. GJ 1243 is one notable example \citep{Davenport_2020}. In the case of \thisstar, the fact that the rotational phase is consistent over both ASAS-SN and \tess\ data shows us that its starspot configuration---and by implication its magnetic field---appears to remain stable for at least six years.

All things considered, despite their different origins and activity levels, it appears that \thisstar\ and a broader class of M dwarfs share with B-type stars the properties that are essential to applications of the RRM model. These similarities cast doubt on the interpretation that a different magnetic field structure is responsible for setting apart centrifugal leakage around high-mass stars and centrifugal breakout around low-mass stars. Accordingly, we now turn to an alternative explanation.

\subsubsection{Stellar Winds or CME?}\label{sec:CME}

For massive stars, the stellar wind is typically considered as the primary source of the plasma trapped in their centrifugal magnetospheres \citep{Townsend_2005, Shultz_2020}, but an M dwarf's stellar wind may not supply enough material for the asymptotic mass to be reached. For low-mass stars, the mass-loss rates from stellar wind  $\dot M_{\rm wind}$ are very difficult to constrain: estimates in the literature differ by nearly five orders of magnitude from $4 \times 10^{-15}$ M$_\odot$  yr$^{-1}$ to 10$^{-10}$ M$_\odot$ yr$^{-1}$ \citep{Vidotto_2013}.

In recent years, it has been suggested that, among young stars, the mass-loss from CMEs $\dot M_{\rm CME}$ may dominate over the outflow from the stellar wind \citep{Jardine_2018}, possibly by one to two orders of magnitude, for solar-mass stars younger than 300 Myr  \citep{Cranmer_2017}. Even more suggestively, \citet{Alvarado_Gomez_2018}'s simulations of low-mass star CMEs indicated that a strong dipolar magnetic field prevents CMEs from breaking free from their stars, albeit with models developed for slowly-rotating stars.
All of this raises the question of whether a different mass-feeding mechanism --- stellar winds for high mass stars, coronal mass ejections for low-mass stars --- could explain the evidence that favors centrifugal leakage for hot stars but centrifugal breakout for low-mass ones.

Moreover, for high-mass stars the magnetosphere would be steadily fed by an isotropic wind, with a high-mass star's perfectly stable fossil magnetic field in the background. This, in turn, would allow the breakout rate to match $\dot M_{\rm wind}$, leading high-mass stars to exhibit the continuous form of centrifugal breakout that \cite{Shultz_2020} introduced as ``centrifugal leakage." Stochastic CMEs, by contrast, would lead to impulsive, localized mass-feeding and the low-mass star's dynamo-generated magnetic field would only be quasi-stable. Accordingly, the asymptotic mass could be reached by the expulsion of a CME, by the gradual feeding of the wind, or by a reorganization of the background magnetic field due to normal dynamo action \cite[\eg][]{Morin_2010}. Under this scenario, centrifugal breakout events could occur suddenly and unpredictably, mirroring the magnetic activity that causes them.

The lack of evidence for centrifugal breakout around high-mass stars offers only circumstantial evidence for the CME mechanism around low-mass stars. Our current data on \thisstar\ is similarly inconclusive: it is consistent with both the CME and stellar wind scenarios. As demonstrated in Section \ref{sec:mass_loss}, the observed dip depth could theoretically be caused by relatively small mass, so that typical M dwarf stellar winds remain a plausible explanation. Future data may more tightly constrain the timescales involved in accumulating the asymptotic mass. Observationally, our LCO data already offers us an upper bound on the timescales involved in the post-breakout reappearance of the dip, but the 107 days between the last \TESS\ dip and the LCO dip is too large for us to eliminate either the stellar wind or the CME scenario. 

Similarly, as discussed in Section \ref{sec:mass_loss}, the largest eclipse equivalent durations occur in a time period that coincides with the highest flare energy. Given the correlation between flares and coronal mass ejections around the Sun \cite[\eg][]{Youssef_2012}, this could be seen as circumstantial evidence for the CME model, but additional data and further numerical modeling, building on the framework developed by \citet{Alvarado_Gomez_2018}, would be needed to confirm this hypothesis. 

%107.334 days\

% Minimum mass loss rate needed: 10^{-16} M\odot / yr

% Msun = 1.989 * 10**33 # g
% time = 107.334/365.25 # years between breakout events
% Mdip = 6 * 10**16/Msun
% Mdip/time

Future observations of \thisstar\ may offer us a unique way to constrain the mass-accumulation rate and to clarify the relationship between magnetospheric clouds and CMEs.  For example, if the reappearance of the dip is shown to be sudden or to occur over timescales too short to be compatible with even the highest estimates of M dwarf stellar winds, this would suggest a trapped CME as the most likely cause of the dips. Regardless, determining the source of \thisstar's accumulating plasma is an important next step. Considering that direct evidence for extra-solar CMEs is almost nonexistent \citep{Alvarado_Gomez_2019_b}, a trapped CME cause might be especially interesting. However, if stellar winds are the source of the dip, that may allow us to better constrain $\dot M_{\rm wind}$ for young and magnetically active low-mass stars. 

\subsection{In the Context of Young Stars}\label{sec:young}

\thisstar, being significantly brighter than many of its analogs, promises to become a benchmark system for understanding a whole class of stars with transiting magnetospheric clouds---systems potentially ranging from the 1.1 Myr B2 star \sig\ to the 7-10 Myr binary M dwarf \PTFO\ \citep{Bouma20} to the 5-10 Myr transient flux-dip stars \citep{Stauffer_2017, Stauffer_2018,  Zhan_2019, Stauffer_2021}. 

This star adds a new name to the short list of flux-dip stars that are candidates for centrifugal breakout, and at only $\approx$44 pc away compared to $\approx$130 pc for the other known centrifugal breakout candidates it is the best choice for follow-up observations. Moreover, at 45 million years old, \thisstar\ is an older analog to the similar stars in Upper Sco that will allow us to probe a new age range, giving us a better understanding of how centrifugal breakout and magnetospheric clouds work at different evolutionary stages. 

We take \thisstar\ to be a good representative of these other systems, in part because they appear tied together by the following characteristics: 

\begin{enumerate}
    \item \textbf{Synchronously rotating dips} not well-explained by a typical transiting planet, but plausibly explained by magnetospheric clouds.
    \item \textbf{Variable dips} with changing morphology, depth, and duration over relatively short timescales, and typically in a gradual manner.
    \item \textbf{A Lack of an Infrared Excess}, in contrast with the class of dipper stars.
    \item \textbf{Asymmetric and triangular dip profiles} in at least some cases.
    \item \textbf{Youth, and strong rotational signals}, likely due at least in part to the fact that these are correlated with strong magnetic fields.
    \item \textbf{H$\alpha$ emission}, with the shape consistent across all phases.
    \item \textbf{One to two dominant dips}, though potentially with smaller secondary ones present. Although this feature could be simple function of geometric orientation rather than probing different physics, we note that this is an observational criterion which distinguishes \thisstar\ and its closest analogs---\sig, \Rik, and the other transient flux-dip stars---from more distant relatives like the persistent flux-dip stars, which have been observed to have up to four dips \citep{Stauffer_2017}.
    \item \textbf{Orbital period under 6 days}, a feature common to \thisstar, both transient and persistent flux-dip stars, \sig, and \PTFO.
    \item \textbf{Occasional sudden disappearances of dips}, sometimes accompanied by an unusually symmetric flare-like event, in a handful of the stars.
\end{enumerate}

More work is needed to clarify which of these characteristics are essential to this emerging class of stars, and to establish where stars that share some, but not all, of these characteristics belong. For example, \sig\ fits characteristics 1, 3, 5, 6, and 8; but there are two major dips and no noticeable variability in its dips' characteristics over time \citep{Townsend_2013}. Similarly, some of the persistent flux-dip stars have dips that disappear suddenly while accompanied by a flare-like event, but they, like \sig, may have multiple mostly stable dips \citep{Stauffer_2017}. 

\section{Conclusions and Future Work} \label{sec:conclusion}

We have presented our analysis of \thisstar, a 45 Myr M dwarf, that has transit-like dips that change in depth and duration over two sectors of \TESS\ data. Besides more gradual depth and duration variations that occur over $\sim$10 day timescales, we take note of a sudden disappearance of the dip over a $\sim$1 day period---and then see the dip reappear in LCO data about 100 days later. We have shown that this behavior can be explained by a magnetospheric cloud and centrifugal breakout scenario and have found the data to be inconsistent with other potential explanations. This model broadly matches the numerical predictions of centrifugal breakout for a star with an inclination similar to \thisstar\ developed by \citet{Townsend_2008}.

When considered alongside the handful of centrifugal breakout candidates among the known flux-dip stars, we argue that we now have observational evidence that centrifugal breakout plays a role in mass-balancing processes for at least some classes of stars---this despite the fact that previous non-detections of centrifugal breakout led astronomers to consider alternative mass-balancing mechanisms. Moreover, \thisstar\ hints at the possibility of uniting a class of mysterious young stars ranging from \sig\ to \PTFO\ and other known flux-dip stars.  

However, \thisstar\ stands out among its potential analogs. It is older, allowing us a glimpse into a different evolutionary stage for this class of stars. It has a long baseline, with more than two years of data, which allows for a better understanding of the transiting dips. This baseline provides the opportunity to observe what appears to be three different dip origins and one phase change, as well as a gradual fading which may be related to recombination of ionized dust during a period of infrequent flares. And, most importantly, \thisstar\ is the brightest of its class of low-mass stars, making it the best choice for follow-up studies and potentially the archetypal system of this kind. 

In particular, future X-ray observations of \thisstar\ may shed light on the origin and structure of these dips. Simultaneous multi-color time-series photometry, ideally via a southern analog of MuSCAT or MuSCAT2 \citep{narita_muscat_2015,narita_muscat2_2019} given
\thisstar's declination $\delta$ of $-63^\circ$, could give a definitive confirmation of the dips' chromaticity. In-depth comparisons with high-mass stars such as the Main-sequence Radio Pulse emitter CU Vir --- a recent centrifugal breakout candidate from \cite{Das_2021} --- may give us insight into how centrifugal magnetospheres vary with spectral type. And tighter photometric constraints on the timescales involved in the reappearance of the dips could allow us to distinguish between the two mass-accumulation mechanisms, CMEs and stellar winds, while potentially adding to currently scanty evidence of extrasolar CMEs or allowing us to more tightly constrain M dwarf stellar wind rates, which, at the moment, are uncertain to five orders of magnitude \citep{Vidotto_2013}. 

\begin{acknowledgments}

We thank Richard Townsend, Katja Poppenhaeger, Trevor David, Christina Hedges, and an anonymous referee for thought-provoking conversations and insights into the \thisstar\ system which improved the quality of this manuscript. We thank Chris Tinney for helpful discussions on Veloce observing modes. E.K.P. thanks  Nils Palumbo (University of Wisconsin-Madison) for much-appreciated coding advice, the NEarby Worlds and Their Stars (NEWTS) group (PI: Benjamin Montet, UNSW) for welcoming her into their lab, and the Caltech Beans for moral support throughout the research process. 

This paper relies on data from the \tess\ mission, which is funded by the NASA Explorer Program. \tess\ data were obtained from the Mikulski Archive for Space Telescopes (MAST), which is supported in part by the NASA Office of Space Science's grant NNX13AC07G.

This research made use of the SIMBAD database,
operated at CDS, Strasbourg, France and of the Exoplanet Follow-up Observation Program (ExoFOP) website, which is operated by the Caltech, under contract with NASA via the Exoplanet Exploration Program.

This work used data from the European Space Agency (ESA) mission {\it Gaia} (\url{https://www.cosmos.esa.int/gaia}), processed by
the {\it Gaia} Data Processing and Analysis Consortium (DPAC, \url{https://www.cosmos.esa.int/web/gaia/dpac/consortium}). Funding
for the DPAC has been provided by national institutions, in particular
the institutions participating in the {\it Gaia} Multilateral Agreement.

This work makes use of observations from the Las Cumbres Observatory
global telescope network, in particular the Sinistro camera on the
LCO-1m telescope at the South African Astronomical Observatory.
The LCO observations were conducted under the auspices of NOIRLab program NOAO2020B-013 (PI: J.~Hartman).

This research also made use of ASAS-SN data. ASAS-SN is funded by the Gordon and Betty Moore Foundation under the 5 year grant GBMF5490 as well as by the National Science Foundation under grants AST-151592 and AST-1908570, with telescopes hosted by LCO. 

This work was also based in part on data obtained at the Anglo-Australian Telescope under program A/2020B/09. We acknowledge the traditional owners of the land on which the AAT stands, the Gamilaraay people, and pay our respects to elders past and present.

This material is based upon work supported by the National Science Foundation Graduate Research Fellowship Program under Grant No. DGE-1746045.

E.K.P.'s role in this research was funded by Caltech's Summer Undergraduate Research Fellowship (SURF) program, with generous support from Carl F. Braun and Harold and Mary Zirin. 

L.G.B. and J.H. acknowledge support by the TESS GI Program, programs G011103 and G022117, through NASA grants 80NSSC19K0386 and 80NSSC19K1728. J.H.\ acknowledges additional support from NASA grant NNX17AB61G.

\end{acknowledgments}

\facilities{\TESS\, ASAS-SN, LCO (SAAO), Siding Spring Observatory (Veloce), MAST, Simbad}

\software{
    astropy \citep{Astropy18},
    CurveFit \citep{CurveFit}
    eleanor \citep{Feinstein_2019},
    emcee \citep{Foreman-Mackey12},
    exoplanet \citep{exoplanet:exoplanet},
    itertools \citep{van1995python},
    lightkurve \citep{lightkurve},
    Mathematica \citep{Mathematica},
    matplotlib \citep{matplotlib},
    numpy \citep{numpy},
    os,
    pickle \citep{van1995python},
    pymc3 \citep{Salvatier2016},
    scipy \citep{scipy},
    statsmodels \citep{seabold2010statsmodels},
    stella \citep{feinstein20_joss},
    theano \citep{exoplanet:theano},
    transitleastsquares \citep{transitleastsquares},
    uncertainties \citep{Uncertainties}}

%\bibliography{exopapers}

\end{document}